%% file: CMDP_TIT_submit.tex
\documentclass[lettersize,onecolumn]{IEEEtran}
\usepackage[margin=1in]{geometry}

\usepackage{ifpdf}
\ifpdf
    \usepackage[pdftex]{graphicx}
    \graphicspath{{figs/}{matlab/}}   
    \usepackage[update]{epstopdf}
\else
	\usepackage{graphicx}
	\graphicspath{{figs/}{matlab/}}   
\fi

\usepackage{tabularx}
\usepackage{booktabs} 
\usepackage[utf8]{inputenc} 
\usepackage[T1]{fontenc}
\usepackage{url}
\usepackage{ifthen}
\usepackage{amsfonts}
\usepackage{mathrsfs}
\usepackage{cite}
\usepackage[cmex10]{amsmath} 
\usepackage{amssymb,amsthm,bm,dsfont}

\usepackage{enumerate}  
\usepackage{color}
\usepackage{xcolor}
\usepackage{cite}
\usepackage{subcaption}
\allowdisplaybreaks[2]  
\usepackage{multirow}
\usepackage{tikz}
\usepackage{cases}
\usepackage{balance}
\newif\ifEnv

\Envtrue
\ifEnv
    
\else
    \excludecomment{Env}
\fi

\usepackage{optidef}
\usepackage{subfig}

\newtheorem{theorem}{\textbf{Theorem}}

\newtheorem{definition}{\textbf{Definition}}

\def\rva{{\mathbf{a}}}

\def\rvd{{\mathbf{d}}}
\def\rve{{\mathbf{e}}}

\def\rvp{{\mathbf{p}}}
\def\rvq{{\mathbf{q}}}

\def\rvs{{\mathbf{s}}}

\def\rvx{{\mathbf{x}}}

\def\rmM{{\mathbf{M}}}

\def\mA{{\bm{A}}}

\def\mD{{\bm{D}}}

\def\mH{{\bm{H}}}

\def\mM{{\bm{M}}}

\def\mP{{\bm{P}}}

\def\mW{{\bm{W}}}

\newcommand{\Expt}{\mathbb{E}}

\newcommand{\argmax}{\arg\max}

\newcommand{\TV}{\text{TV}}

\newcommand{\sign}{\mathrm{sign}}

\allowdisplaybreaks

\definecolor{bl}{rgb}{.2,.2,.8}

\usepackage{xparse}
\NewDocumentCommand{\ml}{m O{M} O{Q_n^{[M]}}} {\mathcal{L}_{#1}(#2\rightarrow #3)}

\usepackage{amsmath}
\usepackage{graphicx}

\begin{document}

\title{Relatively-Secure LLM-Based Steganography via Constrained Markov Decision Processes for High-Efficiency Embedding} 

\author{Yu-Shin Huang, Chao Tian, \IEEEmembership{Fellow, IEEE}, Krishna Narayanan, \IEEEmembership{Fellow, IEEE}, and Lizhong Zheng, \IEEEmembership{Fellow, IEEE}
\thanks{Y.-S. Huang, C. Tian, and K. Narayanan are with the
Department of Electrical and Computer Engineering,
Texas A\&M University, College Station, TX, USA
(e-mail: yushinh@tamu.edu; chao.tian@ece.tamu.edu;
krn@ece.tamu.edu).}
\thanks{L. Zheng is with the Department of Electrical
Engineering and Computer Science,
Massachusetts Institute of Technology, Cambridge, MA, USA
(e-mail: lizhong@mit.edu).}
\thanks{This work was presented in part at the 2025 IEEE International Symposium on Information Theory (ISIT)\cite{2025huangCMDP}}}

\maketitle


\begin{abstract}

    Linguistic steganography aims to conceal information within natural language text without being detected. An effective steganography approach should encode the secret message bits into a minimal number of language tokens while preserving the natural appearance and fluency of the stego-texts. We present a new framework to enhance the embedding efficiency by framing the steganographic embedding procedure as a Constrained Markov Decision Process (CMDP), which takes into consideration the long-term dependencies instead of merely the immediate effects. We constrain the solution space such that the discounted expected cumulative total variation between the selected conditional probability distribution and the original one given by the LLM is below a threshold. To find the optimal policy, we first show that the functional optimization problem can be simplified to a convex problem with a finite number of variables. For the binary-state model, we give a full characterization of the optimal policy, including closed-form expressions in several regimes, and show that it displays water-filling-like behavior in certain cases. The solution indicates that states with highly deterministic transition distributions are preferable for adjustment in the binary-state case. For larger state spaces, the optimal solution is more complex and explicit solutions are difficult to obtain, but the optimal policies show similar structural behaviors.
\end{abstract}


\section{Introduction}

\begin{figure}[t]
    \centering
    \includegraphics[width=0.45\textwidth]{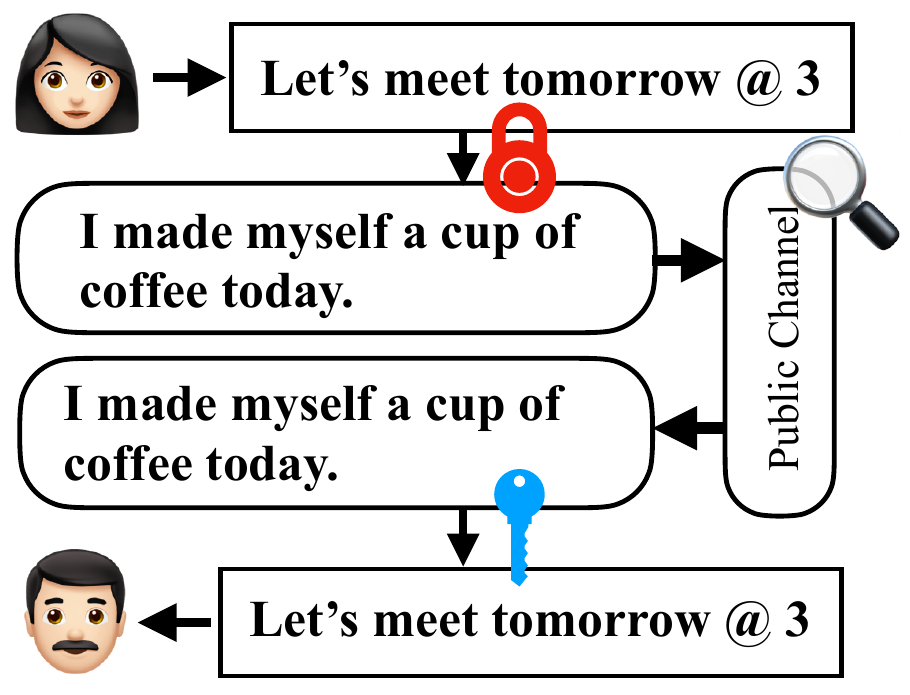}
    \caption{Steganography system\label{fig:stega_explain}}
    \vspace{-0.4cm}
\end{figure}

Steganography allows a sender Alice to communicate a secret message to a receiver Bob by hiding the information inside a carrier signal, such as text, image, audio, or video \cite{anderson1998limits, cox2007digital, provos2003hide}. The encoded signal is transmitted through a public channel observed by an eavesdropper, Eve, whose goal is to determine whether hidden information is being communicated. A classical example is the prisoner's problem \cite{simmons1984prisoners} in which Eve acts as a prison guard who decides whether a letter should be delivered to a prisoner. In this work, we focus on linguistic steganography, where natural language text is used as the carrier signal and the resulting encoded text is referred to as the stego-text; an illustration is shown in Fig. \ref{fig:stega_explain}.

Earlier text steganography approaches mainly hide secret information by modifying an existing cover text \cite{topkara2006hiding, chang2010linguistic, safaka2016matryoshka}. With the development of generative models, especially large language models (LLMs), coverless steganography provides another way to generate the stego-text directly instead of modifying an existing text. Generative steganography methods combine coding techniques such as Huffman coding and arithmetic coding with language models based on RNNs, VAEs, LSTMs, GANs, transformers, and other neural architectures \cite{fang2017generating, yang2018rnn, ziegler2019neural, xiang2017novel, yang2020vae, yang2020gan,de2022perfectly, kaptchuk2021meteor, zhang2024provably, zhang2021provably, ding2023discop}. These approaches can generate natural-looking stego-texts while embedding secret information during the text-generation process. The encrypted secret bits are used as the input to the decompressor, which selects the next token according to the LLM-generated probability distribution. As a result, a probability distribution with higher entropy generally allows more secret message bits to be embedded into each generated token.


In practice, perfect security may not always be necessary, especially when Eve has limited computational capability. Similar considerations have appeared in previous works on ``near-imperceptibility'' steganography \cite{dai2019towards,shen2020near}, where a small amount of deviation is allowed in exchange for other benefits. Following this idea, we consider a relatively-secure steganography setting in which the original LLM-generated probability distribution can be replaced by a nearby distribution when the security requirement is relaxed. Such a replacement provides an opportunity to improve the embedding efficiency. This is particularly useful for LLM-based steganography because several recent works \cite{liao2024co,bai2024semantic,pang2024frestega} have pointed out the low-capacity problem: when the LLM generates a highly concentrated next-token distribution, its entropy becomes small and only a limited number of secret bits can be embedded.

In our recent work \cite{huang2024od}, we studied this relatively-secure steganography problem by optimizing the replacement distribution to maximize its entropy while constraining its deviation from the original LLM-generated distribution. The resulting distribution can increase the amount of secret information embedded into the stego-text. However, this approach treats each token position separately and optimizes only the distribution of the current token. In an autoregressive language model, the token selected at the current step becomes part of the context used to generate later tokens. Therefore, changing the current probability distribution affects not only the embedding efficiency at the current step, but also the probability distributions and embedding efficiency of future tokens. This observation motivates us to consider the effect of the current decision over multiple future token generations.

\begin{figure}[t!]
    \centering
    \includegraphics[width=0.5\textwidth]{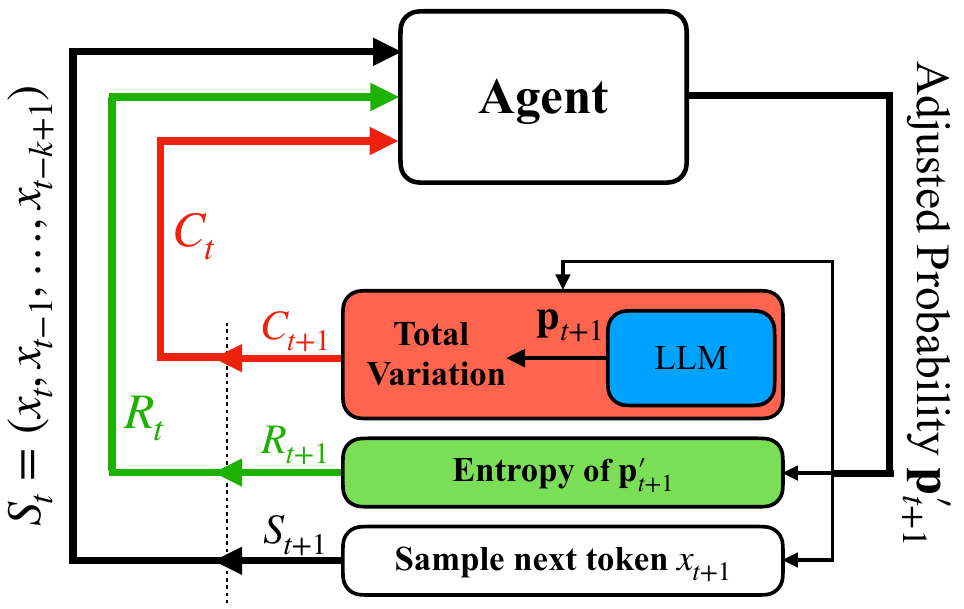}
    \caption{Relatively-secure LLM-based steganography system}
    \label{fig:CMDPstega_explain}
\end{figure}

To take this future effect into consideration, we model the generation process as a decision-making problem. At each step $t$, an agent selects an adjusted probability distribution for generating the next token. The steganography encoder then uses the secret message bits together with this selected distribution to generate the token. Since the encrypted secret bits are i.i.d. random bits, the generated next token is also random. Each selected distribution produces a reward ($R_t$), represented by its entropy and therefore its embedding efficiency, and a cost ($C_t$), represented by its deviation from the original LLM-generated distribution. Meanwhile, the preceding tokens form the state of the system and determine the probability distribution at the next step. As illustrated in Fig. \ref{fig:CMDPstega_explain}, this process naturally fits the framework of a Constrained Markov Decision Process (CMDP) \cite{altman2021constrained}. We therefore formulate relatively-secure LLM-based steganography as a CMDP and optimize its policy under a constraint on the accumulated distribution deviation. 

LLMs are clearly too complex to model precisely. A practical LLM has an extremely large vocabulary and a long context window, making its complete state space too large for direct theoretical analysis. To obtain insights into the structure of the optimal policy, we abstract the LLM into a finite-state model with an $n$-token alphabet and a context length $k$. We first show that the resulting functional CMDP optimization can be reduced to a finite-dimensional convex optimization problem and that an optimal deterministic policy can be obtained from this reduced formulation. We then study three representative models: the binary first-order case $(n,k)=(2,1)$, the binary second-order case $(2,2)$, and the ternary first-order case $(3,1)$. The binary first-order case provides the simplest example and exhibits a water-filling-like solution in some settings. For larger state spaces $(2,2)$, we describe which states and transition probabilities activate as the unnormalized total-variation budget grows, give closed-form optimal policies in the highly constrained regime. We report numerical results for both $(2,2)$ and $(3,1)$ case. These results suggest that adjusting a state with a lower-entropy transition distribution is usually preferred. However, the state to be adjusted cannot be determined from its immediate entropy alone, since the impact on future token generations must also be taken into account.

\section{Preliminaries} \label{sec:prelim}
\subsection{LLM-based Steganography}

A large language model (LLM) outputs a next-token probability distribution $$\rvp_t = [\Pr(X_t = i \mid X_{t-1},X_{t-2},\ldots,X_{t-k})]_{i \in \mathcal{X}}$$ conditioned on the previous tokens within a context window of length $k$, where $X_t$ is the token random variable at time $t$ and $\mathcal{X}$ is the LLM’s token set. By sampling from these distributions autoregressively, an LLM can generate a sequence of natural language tokens \cite{vaswani2017attention,brown2020language,touvron2023llama}. 
We assume that the secret message is already encrypted using a random key shared between Alice and Bob. The encrypted message, denoted by $\mathbf{m}$, can then be modeled as a sequence of i.i.d. uniformly distributed bits.

A key idea in LLM-based steganography is to use the encrypted bit sequence $\mathbf{m}$ to drive the decompressor of a compression algorithm. At time $t$, the decompressor maps the encrypted bits to a token random variable $X_t$ following the LLM-generated distribution $X_t\sim \rvp_t$. The realization $x_t$ is then selected as the next token, which is then included in the context for the LLM to estimate a new probability distribution $\rvp_{t+1}$. Repeating this procedure embeds the secret message while generating the stego-text that is statistically indistinguishable from text without a hidden message.

\begin{figure}[t!]
    \centering
    \includegraphics[width=0.9\textwidth]{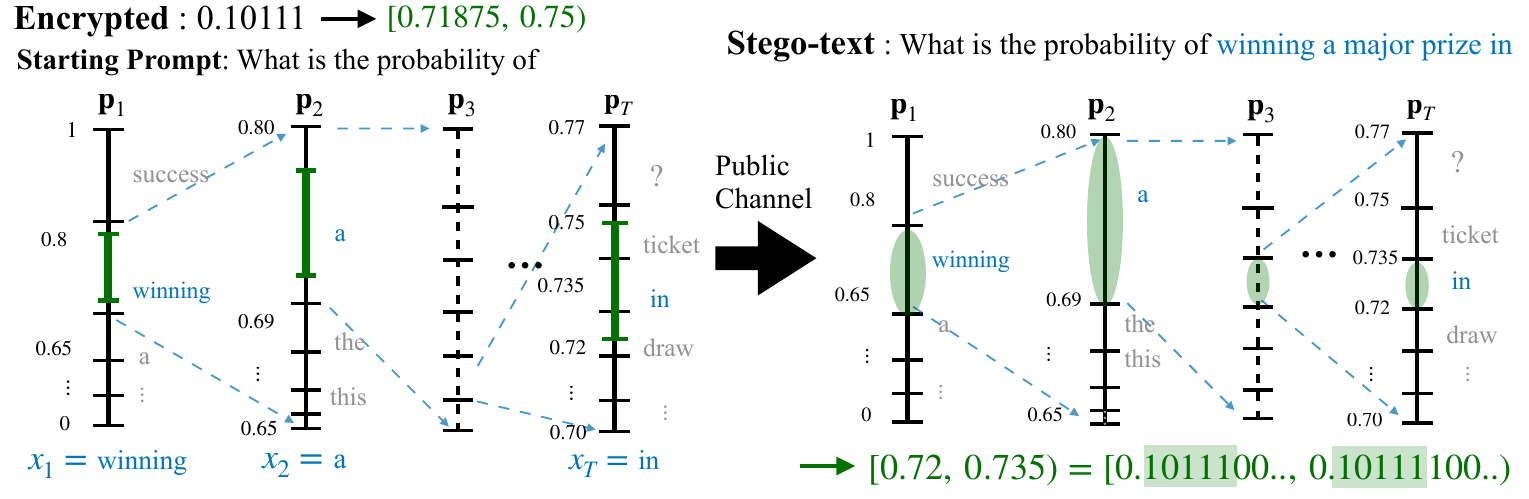}
    \caption{Arithmetic-coding based steganography}\label{fig:AC}
\end{figure}

Arithmetic coding \cite{rissanen1979arithmetic} provides one way to implement this procedure, as illustrated in Fig. \ref{fig:AC}. Consider, for example, an encrypted bit sequence $10111$. This bit sequence represents an interval $$ \mathcal{I} = [0.101110000 \cdots_{2}, 0.1011111111\cdots_{2}) \simeq [0.71875, 0.75).$$ Given a starting prompt, the LLM first produces the next-token probability distribution $\rvp_1$, which partitions the interval $[0,1)$ according to the probabilities of the possible next tokens. 
The interval associated with the encrypted bits specifies the first generated token, \( x_1 = \text{``winning''} \).
Once \( x_1 \) has been determined, the probability distribution \( \rvp_2 \) is obtained by appending \( x_1 \) to the prompt to continue the encoding.

On the receiver's side, Bob divides the received stego-text into tokens and reproduces the same sequence of probability distributions $\{\rvp_t\}$ using the shared LLM and the previous tokens. The arithmetic encoder can then recover the corresponding secret bit sequence. For a stego-text of length $T$, this procedure can embed $\sum_{t=1}^T \Expt [H(\rvp_t)]$ bits of information in expectation. Moreover, since each generated token satisfies $x_t \sim \rvp_t$, the generated token distribution remains the same, resulting in perfectly secure steganography.

\subsection{Embedding Efficiency, Deviation, and State Space Model}
The expected number of secret bits that can be embedded into one generated token is determined by the entropy of its probability distribution. Therefore, when the LLM produces a low-entropy distribution, only a small number of secret bits can be carried by that token. To improve the embedding efficiency, we consider sampling the token from an adjusted distribution $\rvp'_t$ instead of directly using the original LLM-generated distribution $\rvp_t$. At a single time step $t$, this can be viewed as choosing $\rvp'_t$ to maximize its entropy while requiring its deviation from $\rvp_t$ to remain small. This token-wise optimization problem was studied in \cite{huang2024od}.

\begin{figure}[t]
    \centering
    \begin{subfigure}[t]{0.45\linewidth}
        \centering
  \includegraphics[width=\textwidth]{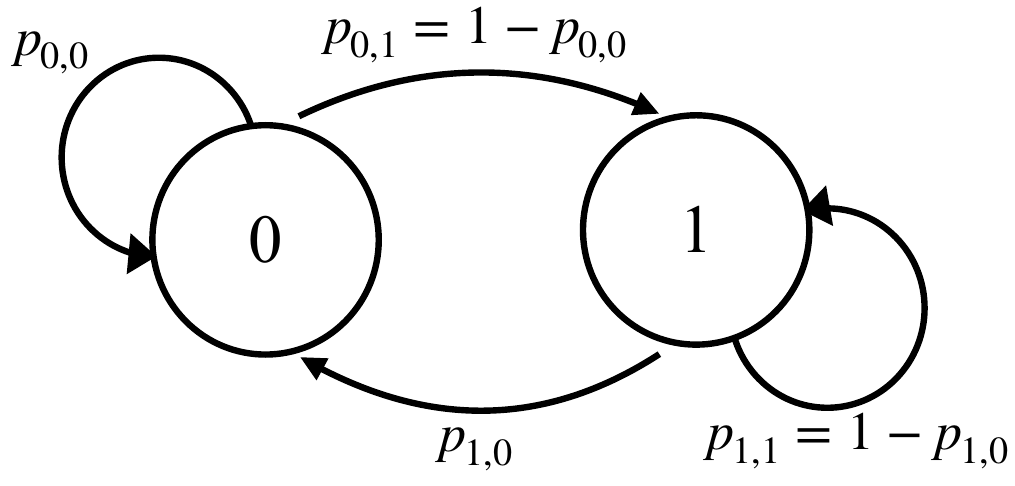}
        \caption{Binary-state model with transition probabilities.}
        \label{fig:two_states}
    \end{subfigure}
    \begin{subfigure}[t]{0.40\linewidth}
        \centering
        \includegraphics[width=\textwidth]{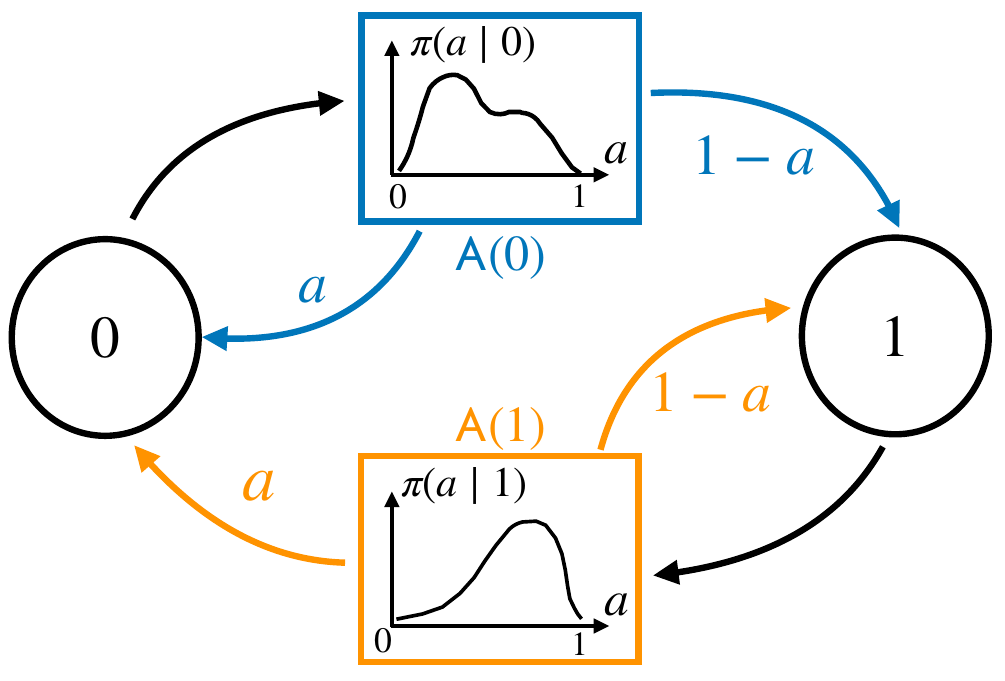}
        \caption{Binary-state CMDP formulation.}
        \label{fig:CMDP2states}
    \end{subfigure}
    \caption{Binary-state language model}
\end{figure}

The autoregressive LLM generation process can also be represented using a state-space model, where each state corresponds to the preceding sequence of $k$ tokens. Suppose that the token alphabet contains $n$ possible tokens. A context of length $k$ then results in $N = n^k$ possible states. For example, when $n=2$ and $k=1$, this language model contains only two states, as illustrated in Fig. \ref{fig:two_states}. In this work, we consider two types of finite-state models. The first uses a context of length $1$, which gives an $n$-state model, while the second uses a context of length $2$, which gives an $n^2$-state model. To keep the theoretical analysis manageable, we focus on the binary first-order, binary second-order, and ternary first-order CMDPs.

\subsection{An Introduction to CMDP} 

A CMDP consists of several key components: the state space $\mathsf{S}$, action space $\mathsf{A}$, transition probabilities $\mP$, reward function $r:\mathsf{S} \times \mathsf{A} \rightarrow \mathbb{R}$, a cost function $g:\mathsf{S} \times \mathsf{A} \rightarrow \mathbb{R}$, a policy $\pi$, the initial state distribution $\mathbf{d}$, and the discount factor $\gamma \in [0,1)$. For each state $s\in\mathsf S$, the policy $\pi$
specifies a probability distribution over the feasible action space
$\mathsf A(s)$.
The goal of policy optimization is to find the optimal policy $\pi$ that guides the agent in selecting actions $a \in \mathsf{A}(s)$ for states $s \in \mathsf{S}$ in order to maximize the cumulative discounted reward over the entire timeframe, i.e.,
\begin{align}
R=(1-\gamma)\Expt\left[\sum_{t=0}^\infty \gamma^t r(S_t,A_t)\right],
\end{align}
subject to the constraint
\begin{align}
C=(1-\gamma)\Expt\left[\sum_{t=0}^\infty \gamma^t g(S_t,A_t)\right]\leq b.
\end{align}
It is often more convenient to describe the CMDP using the discounted
state-action occupancy measure rather than the policy $\pi$. For each
state $s\in\mathsf S$ and measurable set
$\mathcal B\subseteq\mathsf A(s)$, we define
\begin{equation}
    d(s,\mathcal B)
    =
    (1-\gamma)\sum_{t=0}^{\infty}\gamma^t
    \Pr(S_t=s,A_t\in\mathcal B).
\end{equation}
To integrate over the action space, we adopt the measure notation $d(s,\mathrm{d}a)$. Given any function $f(s,a)$, we use
\begin{equation}
    \int_{\mathsf A(s)} f(s,a)\,d(s,\mathrm{d}a)
\end{equation}
to represent the integral of $f(s,a)$ with respect to the occupancy measure $d(s,\cdot)$ in the action coordinate $a \in \mathsf{A}(s)$.
The cumulative reward and cost functions can now be represented in a structured form using $d(s,da)$. 
\begin{align}
    R & = \sum_{s \in \mathsf{S}} \int_{\mathsf{A}(s)} r(s, a) d(s, \mathrm{d}a), \label{eq:CMDP_Ob_def}\\
    C & = \sum_{s \in \mathsf{S}} \int_{\mathsf{A}(s)} g(s,a) d(s,\mathrm{d}a). \label{eq:CMDP_cost_def}
\end{align}

Note that $d(s,\cdot)$ is a non-negative measure, and the discounted occupancy measure satisfies the following occupancy-balance equation:
\begin{align}
    \int_{\mathsf{A}(s)} d(s,\mathrm{d}a) = (1-\gamma)\mathbf{d}_s + \gamma \sum_{s' \in \mathsf{S}} \int_{\mathsf{A}(s')} P(s \mid s', a') \cdot d(s', \mathrm{d}a'). \label{eq:intro_flowconservation}
\end{align}
For a stationary policy, the probability of selecting an action from a
measurable set $\mathcal B\subseteq\mathsf A(s)$ can be obtained from
the occupancy measure as
\begin{equation}
    \pi(\mathcal B\mid s)
    = \frac{\int_{\mathcal B} d(s,\mathrm{d}a)}
    {\int_{\mathsf A(s)}d(s,\mathrm{d}a)} =
    \frac{d(s,\mathcal B)}
    {d(s,\mathsf A(s))},
    \qquad d(s,\mathsf A(s))>0. \label{eq:policy_and_d}
\end{equation}
If \(d(s,\mathsf A(s)) = 0\), the state’s policy \(\pi\) can be selected arbitrarily.
We refer the reader to \cite{sutton2018reinforcement} for more details on MDPs.

\section{Problem Formulation} \label{sec:formulation}

\subsection{State Space, Action Space, and Policies}

\begin{figure}[t!]
    \centering
    \includegraphics[width=0.8\textwidth]{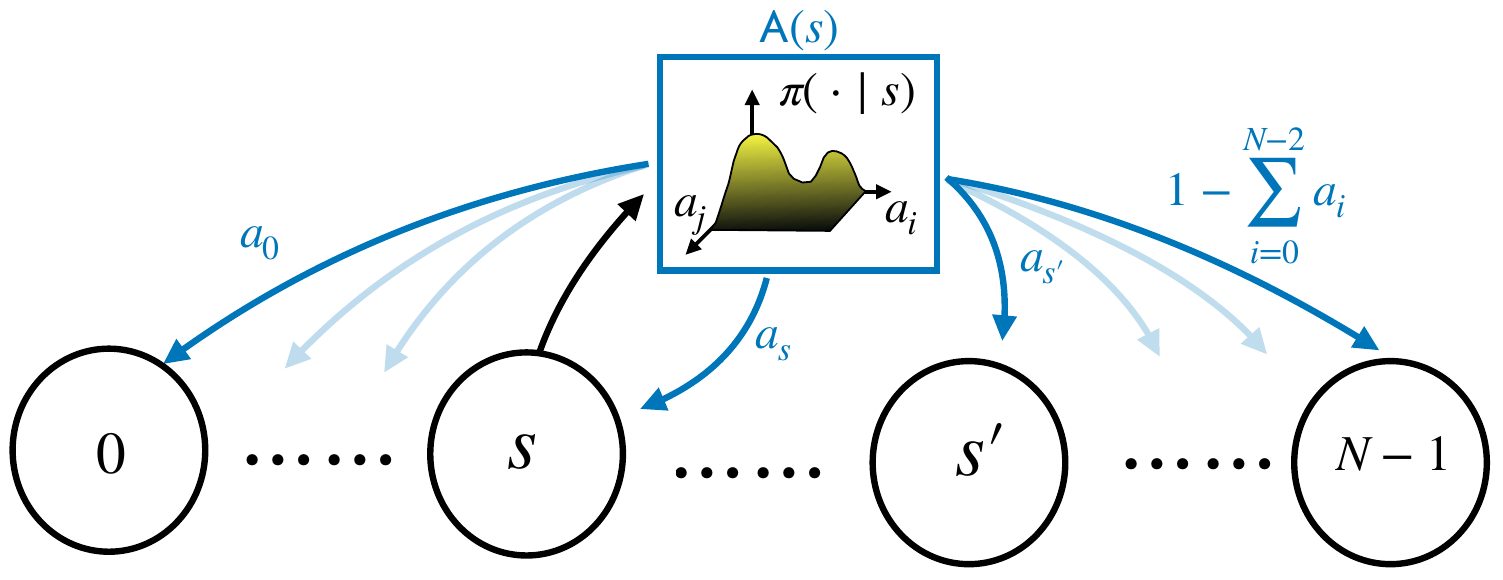} 
    \caption{Steganography with CMDP: $N$-state model} 
    \label{fig:CMDP_n_states}
\end{figure}

To formulate steganographic coding as a CMDP problem, we begin by defining the relevant components as described above. Figure \ref{fig:CMDP2states} shows the binary-state transitions of the CMDP, whereas Figure \ref{fig:CMDP_n_states} presents the CMDP with $N$-state transitions, focusing specifically on state $s$.

As described above, we work with an abstract state space formed by an $n$-token vocabulary and the preceding context of length $k$. The token set is $\mathcal{X} = \{0,1,\dots,n-1\}$, and the state space is $\mathsf{S} = \mathcal{X}^k$. Thus, the state space $\mathsf{S}$ contains $N = n^k$ distinct states and the state at time $t$ can be written as $S_t = (X_t, X_{t-1}, \dots, X_{t-k+1}) \in \mathsf{S}$. A generic state $\rvs \in \mathsf{S}$ can be written as $\rvs = (s^{(1)}, s^{(2)}, \dots, s^{(k)})$, where $s^{(1)}$ is the most recent token. We then define the set of admissible successor states by
\begin{align}
    \Gamma(\rvs) = \{(i, s^{(1)}, s^{(2)}, \dots, s^{(k-1)}): i \in \mathcal{X} \},
\end{align}
which lists all possible states that can immediately follow the state $\rvs$. 

We denote 
\begin{align}
    p_{\rvs,i} = \Pr(X_{t+1} = i \mid S_t = \rvs), \label{eq:def_p_s,i}
\end{align}
as the conditional probability of generating token $i$ from state $\rvs$.
For each state $\rvs$, we then define the transition probabilities by $\rvp_{\rvs} = [p_{\rvs,\rvs^{\prime }}]_{\rvs^{\prime } \in \mathsf{S}}$, where
\begin{align}
    p_{\rvs,\rvs^{\prime }}=\begin{cases}
        p_{\rvs,i}, & \text{if }\rvs^{\prime }=(i, s^{(1)}, s^{(2)}, \dots, s^{(k-1)}),\ i \in \mathcal{X},\\[4pt]
        0, & \text{if }\rvs^{\prime }\notin \Gamma(\rvs)
    \end{cases}.
\end{align}
In this setting, we specify an action space associated with state $\rvs$ as
\begin{align}
    \mathsf{A}(\rvs) = \{ \rvq \in \Delta_N : \rvq_{\rvs'} = 0 \text{ for all } \rvs' \notin \Gamma(\rvs) \}, 
\end{align}
where $\Delta_N := \{ [q_{\rvs'}]_{{\rvs'}\in \mathsf{S}} \in [0,1]^N: \sum_{\rvs' \in \mathsf{S}} q_{\rvs'} = 1\} $. An action selected at state $\rvs$ is denoted by
\begin{align}
    \rva = [a_{\rvs'}]_{\rvs'\in\mathsf{S}} \in \mathsf{A}(\rvs),
\end{align}
where $a_{\rvs'}$ denotes the probability assigned to the successor
state $\rvs'$. Thus, choosing action $\rva$ induces the transition
probability
\begin{align}
    P_{\rva}(\rvs,\rvs')
    &\triangleq
    \Pr\!\left(
        S_{t+1}=\rvs'
        \mid
        S_t=\rvs,
        A_t=\rva
    \right)
    = a_{\rvs'}.
\end{align}
For simplicity, we write the transition probability vector induced by
$\rva$ at state $\rvs$ as
\begin{align}
    [P_{\rva}(\rvs,\rvs')]_{\rvs'\in\mathsf{S}}
    =
    [a_{\rvs'}]_{\rvs'\in\mathsf{S}}
    =
    \rva.
\end{align}
Since $\rva$ is the transition probability vector induced by the selected
action, the reward function is
\begin{align}
    r(\rvs,\rva)
    = H([P_{\rva}(\rvs,\rvs')]_{\rvs'\in\mathsf{S}})= 
    H(\rva),
    \qquad
    \rvs\in\mathsf{S},\;
    \rva\in\mathsf{A}(\rvs).
\end{align}
where $H(\cdot)$ is the Shannon entropy function with base $2$. Throughout this paper, we assume a discount factor $\gamma$ with $0<\gamma<1$. If $\gamma=0$, the model ignores future effects, and the resulting formulation matches our previous work \cite{huang2024od}.  
The selected transition distribution $[P_{\rva}(\rvs,\rvs')]_{\rvs'\in\mathsf{S}}$ should remain close to the
original transition distribution $\mathbf p_\rvs$. Accordingly, we define
the constraint cost function as
\begin{align}
    g(\rvs,\rva)
    = \overline{\TV}([P_{\rva}(\rvs,\rvs')]_{\rvs'\in\mathsf{S}},\mathbf p_\rvs) = 
    \overline{\TV}(\rva,\mathbf p_\rvs),
    \qquad
    \rvs\in\mathsf{S},\;
    \rva\in\mathsf{A}(\rvs).
    \label{eq:cost_TV}
\end{align}
where, and throughout this work, we quantify the deviation using the unnormalized total variation distance
\begin{align}
\overline{\TV}(\rva, \mathbf{p}_\rvs) = \| \rva - \rvp_\rvs \|_1 = \sum_{\rvs' \in \mathsf{S}}|a_{\rvs'} - p_{\rvs,\rvs'}|,
\end{align}
which equals twice the standard (normalized) total variation distance.
In the special case where $(n = 2, k = 1)$, the unnormalized total variation distance becomes
\begin{align}
\overline{\TV}(\rva, \mathbf{p}_\rvs) &= |a_0 - p_{\rvs,0}| + |a_1 - p_{\rvs,1}| \notag \\
&= |a_0 - p_{\rvs,0}| + |1 - a_0 - (1 - p_{\rvs,0})| \notag \\
&= 2|a_0 - p_{\rvs,0}|, \label{eq:TV_budget_derive_n=2}
\end{align}
which is equal to $2$ times the distance between the first components.

\subsection{CMDP Formulation}

With the components given above, we can utilize the general CMDP
framework and formulate our problem in terms of the discounted
occupancy measures $d(s,\cdot)$, denoted as ($\mathbf{P1}$) below. 
\begin{align}
    \mathbf{P1}: ~~~\max_{\{d(\rvs,\cdot)\}_{\rvs \in \mathsf{S}}}: & ~ \sum_{\rvs \in \mathsf{S}} \int_{
    \mathsf{A}(\rvs)}  H(\rva) d(\rvs,\mathrm d\rva) \label{eq:opt_problem_ob_TV} \\
    \text{subject to: } & ~ \int_{\mathsf{A}(\rvs)} d(\rvs,\mathrm d\rva) = \rvd^{\gamma}_\rvs+\gamma \sum_{\rvs'\in \mathsf{S}} \int_{\mathsf{A}(\rvs')} a_{\rvs} d(\rvs',\mathrm d\rva), \; \forall \rvs \in \mathsf{S} \label{eq:con_same1_TV}\\
    & ~ \sum_{\rvs \in \mathsf{S}} \int_{\mathsf{A}(\rvs)}\overline{\TV}(\rva, \mathbf{p}_\rvs) d(\rvs,\mathrm d\rva) \leq b \label{eq:opt_problem_con3_TV}\\
    & ~ d(\rvs,\cdot) \text{ is a non-negative measure }, ~ \forall \rvs \in \mathsf{S}\label{eq:opt_problem_con4_TV}
\end{align}
where we denote the discounted initial probabilities $(1-\gamma)\mathbf{d}_{\rvs}$ as $\mathbf{d}^{\gamma}_\rvs$.

Once the optimal $d(\rvs,\cdot)$ is obtained from problem $\mathbf{P1}$, the optimal policy $\pi$ can be derived from the associated conditional occupancy distribution, in the same manner as in (\ref{eq:policy_and_d}). Since the decision
variable $d(\rvs,\cdot)$ is a measure over the continuous action space
$\mathsf A(\rvs)$ for each state $\rvs$, problem $\mathbf{P1}$ is an
infinite-dimensional optimization problem, even though its objective
and constraints are linear in the occupancy measures.

\subsection{The Reduced Formulation} \label{sec:simplfy_opt}
In this section, we show that the infinite-dimensional optimization problem \(\mathbf{P1}\) can be reformulated as the finite-dimensional optimization problem \(\mathbf{P3}\), which involves \(N^2+N\) scalar decision variables. Mathematically, $\mathbf{P3}$ is a standard finite-dimensional convex optimization problem, which is significantly simpler than the functional optimization problem $\mathbf{P1}$. Operationally, once we solve $\mathbf{P3}$, we can obtain the parameters $a_{\rvs,\rvs'}, \forall \rvs,\rvs' \in \mathsf{S}$ that immediately give us the optimal policy: at state $\rvs$, choose the probability distribution $\rva_s = [a_{\rvs,\rvs'}]_{\rvs' \in \mathsf{S}}$ to generate the next token (state). The policy itself is deterministic, and the randomness in generating the next token comes purely from the encrypted message bits.

Let us begin by defining a new optimization problem $\mathbf{P2}$ as follows. 
\begin{align}
    \mathbf{P2}: ~\max_{
        \substack{(\rva_\rvs , d_\rvs)_{\rvs \in \mathsf{S}} \\
        \rva_\rvs = [a_{\rvs, \rvs'}]_{\rvs' \in \mathsf{S}}}}: & ~ \sum_{\rvs\in \mathsf{S}} d_\rvs H(\rva_\rvs) \\
    \text{subject to:} & ~  d_\rvs = \rvd^\gamma_\rvs + \gamma \sum_{\rvs' \in \mathsf{S}} a_{\rvs',\rvs}d_{\rvs'},  \forall \rvs \in \mathsf{S} \\
    & ~ \sum_{\rvs \in \mathsf{S}} d_\rvs \overline{\TV}_\rvs(\rva_\rvs) \leq b \\
    & ~ d_\rvs \geq 0, \forall \rvs \in \mathsf{S}
    \\
    &~ 0\le a_{\rvs,\rvs'} \le 1,  \forall \rvs,\rvs'\in\mathsf S,
    \text{ and } \sum_{\rvs'\in\mathsf S}a_{\rvs,\rvs'}=1, \forall \rvs \in\mathsf S,
    \\
    & ~ a_{\rvs,\rvs'}=0, ~
    \rvs'\notin\Gamma(\rvs), ~\forall \rvs\in\mathsf S. \label{eq:p2_constraint_a0a1}
\end{align}
For simplicity, we represent  $\overline{\TV}({\rva_\rvs}, \mathbf{p}_\rvs)$ with $\overline{\TV}_\rvs(\rva_\rvs)$. 
\begin{theorem}\label{thm:reduce_varto2}
 $(\mathbf{P1})=(\mathbf{P2})$, where ``$=$`` stands for equivalence between the two problems. 
    \end{theorem}
The proof of Theorem \ref{thm:reduce_varto2} is detailed in Appendix A 
, where we utilize the concavity of the objective function and the convexity of  constraint functions to show that a weighted average of the probabilistic policy is without loss of optimality. 
Since $\mathbf{P2}$ is not necessarily convex, we next show that a reparametrization of the problem leads to another equivalent problem $\mathbf{P3}$ that is indeed convex. 
\begin{align}
    \mathbf{P3}: ~ \max_{
        \substack{(\rvx_\rvs , d_\rvs)_{\rvs \in \mathsf{S}} \\
        \rvx_\rvs = [x_{\rvs, \rvs'}]_{\rvs' \in \mathsf{S}}}}:   & \; \; \sum_{\rvs \in \mathsf{S}} d_\rvs H(\tfrac{\rvx_\rvs}{d_\rvs}) \label{eq:x0_obj_func_TV} \\
    \text{subject to:} & ~  d_\rvs = \rvd^\gamma_\rvs + \gamma \sum_{\rvs' \in \mathsf{S}} x_{\rvs',\rvs}, \forall \rvs \in  \mathsf{S} \label{eq:x0_constraint_1_TV}\\
     & ~ \sum_{\rvs \in \mathsf{S}} d_\rvs \overline{\TV}_\rvs(\tfrac{\rvx_\rvs}{d_\rvs}) \leq b \label{eq:x0_constraint_3_TV}\\
    & ~ 0 \leq x_{\rvs,\rvs'} \leq d_\rvs , \forall \rvs, \rvs'  \in \mathsf{S} \text{ and } \sum_{\rvs' \in \mathsf{S}} x_{\rvs,\rvs'} = d_\rvs, ~ \forall \rvs \in \mathsf{S},\label{eq:x0_constraint_5_TV} \\
    & x_{\rvs,\rvs'} = 0 \text{ if } \rvs' \notin \Gamma(\rvs), \forall \rvs \in \mathsf{S} \label{eq:x0_constraint_6_TV}
\end{align}
The number of variables in $\mathbf{P3}$ is now $N^2+N$. Clearly, we have substituted $\rva_\rvs$ in $\mathbf{P2}$ with $\rvx_\rvs = \rva_\rvs d_\rvs$, and the two problems are equivalent. Because constraints (\ref{eq:x0_constraint_1_TV}),(\ref{eq:x0_constraint_5_TV}), and (\ref{eq:x0_constraint_6_TV}) are linear, it remains only to verify whether the objective function (\ref{eq:x0_obj_func_TV}) and the constraint (\ref{eq:x0_constraint_3_TV}) are concave and convex, respectively. Recall that the perspective function $f_2(x,t)$ of a function $f_1(x)$ is defined as 
\begin{align}
    f_2(x,t) = t f_1\left(\frac{x}{t}\right), ~ \forall t > 0
\end{align}
A well-known property of the perspective function is that if $f_1(\cdot)$ is convex/concave, $f_2(\cdot,\cdot)$ is likewise convex/concave. Since the entropy function $H(\cdot)$ is concave and the unnormalized total variation $\overline{\TV}$ function is convex, (\ref{eq:x0_obj_func_TV}) is indeed concave and (\ref{eq:x0_constraint_3_TV}) is indeed convex, implying that the problem $\mathbf{P3}$ is a convex optimization problem. When replacing $\rva_\rvs$ by $\frac{\rvx_\rvs}{d_\rvs}$, we must guarantee that this expression remains well defined in the case $d_\rvs=0$. If $d_\rvs=0$, then the constraint \eqref{eq:x0_constraint_5_TV} enforces $x_{\rvs,\rvs'} = 0$. Consequently, we set
the associated perspective terms at $(\rvx_\rvs,d_\rvs)=(\mathbf{0},0)$ as
\[
d_\rvs H\!\left(\tfrac{\rvx_\rvs}{d_\rvs}\right)=0,
\qquad
d_\rvs \overline{\TV}_\rvs\!\left(\tfrac{\rvx_\rvs}{d_\rvs}\right)=0.
\]

In summary, standard discounted CMDP theory converts the policy optimization problem into the occupation measure formulation $\mathbf{P1}$, but because the action spaces are continuous simplices, $\mathbf{P1}$ remains infinite-dimensional. Exploiting the convexity of the action spaces, the affine dependence of the transition kernels on the actions, the concavity of entropy, and the convexity of total variation, we show that every randomized stationary policy is dominated by a deterministic stationary policy. This reduces $\mathbf{P1}$ to the finite-dimensional problem $\mathbf{P2}$. A perspective reparameterization then yields the convex formulation $\mathbf{P3}$.

\section{Optimal Policy for the Binary First-Order CMDP}\label{sec:results}

We derive the optimal policy $\pi$ by solving problem $\mathbf{P3}$ in the specific case $(n=2,k=1)$. In this binary-state setting, the state space reduces to
$\mathsf{S}=\{0,1\}$. Since each state is represented by a single token index
rather than a vector, we use $s$ instead of $\rvs$ to denote the state in this
section. For each $s\in\mathsf{S}$, both states are possible successor states,
and therefore $\Gamma(s)=\mathsf{S}$. Consequently, the action space is
$\mathsf{A}(s)=\Delta_2$, and the action at state $s$ can be parameterized as $\rva_s=[a_{s,0},\,1-a_{s,0}] \in \mathsf{A}(s)$.

$\mathbf{P3}$ can be further reduced to the problem $\mathbf{P3\text{-}(2,1)}$. While $\mathbf{P3}$ originally involves $6$ variables, the number of decision variables in $\mathbf{P3\text{-}(2,1)}$ decreases to $4$ by substituting $x_{s,1} = d_s - x_{s,0}$.
\begin{align}
    \mathbf{P3\text{-}(2,1)}: ~ \max_{d_0,d_1, x_{0,0},x_{1,0}}  & \; \; d_0 H\!\left(\tfrac{x_{0,0}}{d_0}\right) + d_1 H\!\left(\tfrac{x_{1,0}}{d_1}\right) \label{eq:x0_obj_func_TV_2s} \\
    \text{subject to:} & ~  d_0 = \rvd^\gamma_0 + \gamma (x_{0,0} + x_{1,0}) \label{eq:x0_constraint_1_TV_2s}\\
    & ~ d_1 = \rvd^\gamma_1 + \gamma (d_0 -x_{0,0} + d_1 - x_{1,0})
    \label{eq:x0_constraint_2_TV_2s} \\
    & ~ d_0 \Bigl|\tfrac{x_{0,0}}{d_0}-p_{0,0}\Bigr|+d_1 \Bigl|\tfrac{x_{1,0}}{d_1}-p_{1,0}\Bigr| \leq \tfrac{b}{2} \label{eq:x0_constraint_3_TV_2s}\\
    & ~ 0 \leq x_{s,0} \leq d_s , \forall s \in \{ 0,1\} \label{eq:x0_constraint_5_TV_2s}
\end{align}
In $\mathbf{P3\text{-}(2,1)}$, $H(\cdot)$ reduces to the binary entropy function. The TV budget constraint \eqref{eq:x0_constraint_3_TV_2s} is obtained from \eqref{eq:TV_budget_derive_n=2}. Moreover, the vector $\rvx_s$ takes the form $\rvx_s = [x_{s,0}, d_s-x_{s,0}]$, and the action of each state is given by $\rva_s = \frac{\rvx_s}{d_s} = [a_{s,0}, 1-a_{s,0}]$. By summing constraints \eqref{eq:x0_constraint_1_TV_2s} and \eqref{eq:x0_constraint_2_TV_2s}, we obtain
\begin{align}
    d_0 + d_1 = 1. \label{eq:x0_constraint_2_TV_2s_sum1} 
\end{align}
Since the constraint pair \eqref{eq:x0_constraint_1_TV_2s}, \eqref{eq:x0_constraint_2_TV_2s} is equivalent to the pair \eqref{eq:x0_constraint_1_TV_2s}, \eqref{eq:x0_constraint_2_TV_2s_sum1}, we shall replace  \eqref{eq:x0_constraint_2_TV_2s} with \eqref{eq:x0_constraint_2_TV_2s_sum1} when considering this problem. Moreover, since the optimal action $\rva_s$ varies with the budget $b$, we may denote it as $\rva_s(b)$. We further introduce the notion of an active state to more clearly characterize our solution to this problem.
\begin{definition}[Active state] \label{def:active-state}
    Given an unnormalized total variation budget $b$ for any instance of problem $\mathbf{P3}$, we say that $s$ is in an active state when the optimal action is not the same as the transmission probability, i.e., $\rva_\rvs(b) \neq \rvp_\rvs$. 
\end{definition}

\subsection{Solution Regimes for $\mathbf{P3\text{-}(2,1)}$}

In this binary-state CMDP, the solutions fall into three regimes based on the value of the unnormalized total variation budget $b$: 
\begin{enumerate}
\item Regime I: Highly constrained regime \\
A single active state; In the state that produces lower-entropy tokens, the distribution becomes more uniform, while in the other state the original distribution is preserved; 
\item Regime II: Intermediately constrained regime \\
Both states become active; In both states $0$ and $1$, the original distributions become more uniform;
\item Regime III: Unconstrained regime \\
In both states, the distributions become fully uniform, namely $\rva_s = [\tfrac{1}{2}, \tfrac{1}{2}] $.
\end{enumerate}

\subsection{Solution to $\mathbf{P3\text{-}(2,1)}$}
Throughout this analysis we assume $0<p_{s,0}<1$ for each $s\in\{0,1\}$, ensuring that the transition probabilities are nondegenerate. We then focus on the regimes $p_{0,0},p_{1,0}\geq \frac{1}{2}$, $p_{1,0} \geq \frac{1}{2}>p_{0,0}$, and $p_{0,0} \geq \frac{1}{2}>p_{1,0}$. When $p_{0,0},p_{1,0}\leq \frac{1}{2}$, the same solution follows by renaming the two states; see Appendix C
for details.  
Let us also define the following auxiliary functions:
   \begin{align}
            \eta_0^{\pm}(b) & = \frac{\frac{b}{2} (1+\gamma p_{1,0})\pm p_{0,0}(-\mathbf{d}^{\gamma}_0 -\gamma p_{1,0})}{\pm(-\gamma p_{1,0}-\mathbf{d}^{\gamma}_0) +\gamma \frac{b}{2} } \label{eq:eta_0_def}\\
            \eta_1^{\pm}(b) & = \frac{\frac{b}{2} (1-\gamma p_{0,0})\pm p_{1,0}(1-\gamma p_{0,0}-\mathbf{d}^{\gamma}_0 )}{\pm (1-\gamma p_{0,0}-\mathbf{d}^{\gamma}_0) -\gamma \frac{b}{2} }. 
    \end{align}
The $\pm$ superscripts on $\eta_0$ and $\eta_1$ specify which sign to take wherever $\pm$ appears in the formula. For instance, $\eta_0^-$ means that in \eqref{eq:eta_0_def} one should consistently choose the $-$ option whenever a $\pm$ occurs.
The following two theorems give the precise optimal policy, the proofs of which can be found in Appendix A. 
\begin{theorem} \label{thm:p0p1sameside}
    When $p_{0,0}, p_{1,0} \geq \frac{1}{2}$, the optimal policy choices at the two states are:
    \begin{enumerate}
        \item Regime I: $b \in [0, b_1) $ \label{thm:2-1}
        \begin{align}
            \begin{cases}
                \rva_0 = \rvp_0, ~ \rva_1 = [\eta_1^{-}(b), 1-\eta_1^{-}(b)] , &\mbox{if}~ |\frac{1}{2} - p_{1,0} | \geq |\frac{1}{2} - p_{0,0} |\\
                \rva_0 = [\eta_0^{+}(b), 1-\eta_0^{+}(b)] , ~ \rva_1 = \rvp_1, &\mbox{if} ~ |\frac{1}{2} - p_{1,0} | < |\frac{1}{2} - p_{0,0} | 
            \end{cases}
        \end{align}
        \item Regime II: $b \in [b_1, b_2)$ \label{thm:2-2}
        \begin{align}
            & \rva_0 = \rva_1 = [-M(\tfrac{b}{2} - \mathbf{d}_{0}^\gamma p_{0,0} - \mathbf{d}_{1}^\gamma p_{1,0} -\gamma p_{1,0}), 1 + M(\tfrac{b}{2} - \mathbf{d}_{0}^\gamma p_{0,0} - \mathbf{d}_{1}^\gamma p_{1,0}-\gamma p_{1,0} ) ] 
        \end{align}
        \item Regime III: $b \in [b_2, \infty)$ \label{thm:2-3}
        \begin{align}
            & \rva_0 = \rva_1 = [\tfrac{1}{2}, \tfrac{1}{2}]
        \end{align}
    \end{enumerate}
    where 
    \begin{align}        
            b_1 & = 2\max\{(1 -\gamma p_{0,0} - \mathbf{d}^{\gamma}_0 )(p_{1,0} - p_{0,0}),(\mathbf{d}^{\gamma}_0 +\gamma p_{1,0})(p_{0,0} - p_{1,0})\}\\
		b_2 & = (2\mathbf{d}^{\gamma}_0 +\gamma)(p_{0,0} - p_{1,0})+2p_{1,0}-1\\
        M & = \frac{1}{1-\gamma p_{0,0} + \gamma p_{1,0}}.
    \end{align}    
\end{theorem}

\begin{theorem}\label{thm:p0p1diffside}
    If $p_{1,0}$ and $p_{0,0}$ lie on opposite sides of $\frac{1}{2}$, the optimal policy decisions in the two states are:
    \begin{enumerate}
        \item Regime I: $b \in [0, b_1^{\diamond}) $
        \begin{align}
            \begin{cases}
                \rva_0 = \rvp_0, ~ \rva_1 = [\eta_1^{\diamond}(b), 1-\eta_1^{\diamond}(b)] &, \mbox{if}~ |\frac{1}{2} - p_{1,0} | \geq |\frac{1}{2} - p_{0,0} |\\
                \rva_0 = [\eta_0^{\diamond}(b), 1-\eta_0^{\diamond}(b)] , ~ \rva_1 = \rvp_1 &, \mbox{if} ~ |\frac{1}{2} - p_{1,0} | < |\frac{1}{2} - p_{0,0} | 
            \end{cases}
        \end{align}
        \item Regime II: $b \in [b_1^{\diamond}, b_2^{\diamond}]$
        \begin{align}
            &(a_{0,0}, a_{1,0}) \in S_b^{\diamond}  = \bigg\{ (a_{0,0},a_{1,0}) \in \mathcal{I}_0 \times \mathcal{I}_1 ~ :~\nonumber 
            \\ &\qquad \Phi_-(a_{1,0}) = \Phi_+(a_{0,0}) 
             \text{ and } m^{\diamond}(a_{0,0},a_{1,0}) = \tfrac{b}{2} \bigg\} \label{eq:thm3_regimeII_sol}
        \end{align}
        \item Regime III: $b \in (b_2^{\diamond}, \infty)$ 
        \begin{align}
            & \rva_0 = \rva_1 =[ \tfrac{1}{2}, \tfrac{1}{2}]
        \end{align}
    \end{enumerate}
    where  
    \begin{align} 
            \diamond & = \begin{cases}
                +, & \text{if} \quad p_{0,0} \geq \frac{1}{2} \geq p_{1,0} \\
                -, & \text{if} \quad p_{1,0} \geq \frac{1}{2} \geq p_{0,0}
            \end{cases}, \quad \mathcal{I}_s = \left[ \min(p_{s,0}, \frac{1}{2}), \max(p_{s,0}, \frac{1}{2})\right], \forall s \in \{ 0,1\}\\
            b_1^{\pm}  & = 
            \begin{cases}
                \pm2\left( \tfrac{\mathbf{d}^{\gamma}_0 +\gamma p_{1,0}}{1-\gamma \psi_{0} +\gamma p_{1,0}} \right)  (p_{0,0} -\psi_{0} ),  & \text{if } |\tfrac{1}{2} - p_{1,0}| < |\tfrac{1}{2} - p_{0,0}| \\
                \pm 2\left( \tfrac{1-\gamma p_{0,0}-\mathbf{d}^{\gamma}_0 }{1-\gamma p_{0,0}+\gamma \psi_{1} } \right)  (\psi_{1}-p_{1,0}), & \text{otherwise}
            \end{cases}  \label{eq:bl'_def} \\
		b_2^{\pm}  & = \pm(2\mathbf{d}^{\gamma}_0 +\gamma)(p_{0,0} + p_{1,0}-1)\pm (1 -2p_{1,0}) \label{eq:bh'_def}
        \end{align}
    and
        \begin{align}        
            \Phi_{\pm}(a) & = ( \pm1 + \gamma p_{0,0} + \gamma p_{1,0}) \log \left(\frac{1-a}{a}\right) -2 \gamma \log(1-a) \\
         \psi_1 & \in \{ a \in \mathcal{I}_1 \mid \Phi_+(p_{0,0}) = \Phi_-(a)\}, \\
        \psi_0 & \in \{ a \in \mathcal{I}_0 \mid \Phi_-(p_{1,0}) = \Phi_+(a) \}  \\
        m^{\pm}(a_{0,0},a_{1,0}) & = \pm d_0\left( p_{0,0}-a_{0,0} \right)\pm d_1\left( a_{1,0}-p_{1,0} \right)        
    \end{align}
    where $u(\cdot)$ denotes the unit-step function. 
\end{theorem}

The two theorems above characterize the optimal action solution for the binary first-order setting. Theorem \ref{thm:p0p1sameside} gives a closed-form solution for all regimes, while Theorem \ref{thm:p0p1diffside} provides closed-form expressions for regimes I and III. For regime II, the proof in Appendix B
establishes that, for any $b$ in this regime, the solution is unique, so $\rva_0$ and $\rva_1$ can be computed numerically but a closed-form solution appears difficult. We further note that the state visitation frequency $d_s$ can be expressed in terms of $a_{0,0}$ and $a_{1,0}$: 
\begin{align}
    d_0 & = \frac{\mathbf{d}^{\gamma}_0 + \gamma a_{1,0}}{1 - \gamma a_{0,0} + \gamma a_{1,0}} , \quad d_1 = \frac{1 - \gamma a_{0,0}-\mathbf{d}^{\gamma}_0}{1 - \gamma a_{0,0} + \gamma a_{1,0}} \label{eq:dsvalue}
\end{align}

\subsection{Numerical Results for Binary First-Order CMDP}

\begin{figure}[t]
    \centering
    \begin{subfigure}[b]{0.32\textwidth}
        \centering
        \includegraphics[width=\textwidth]{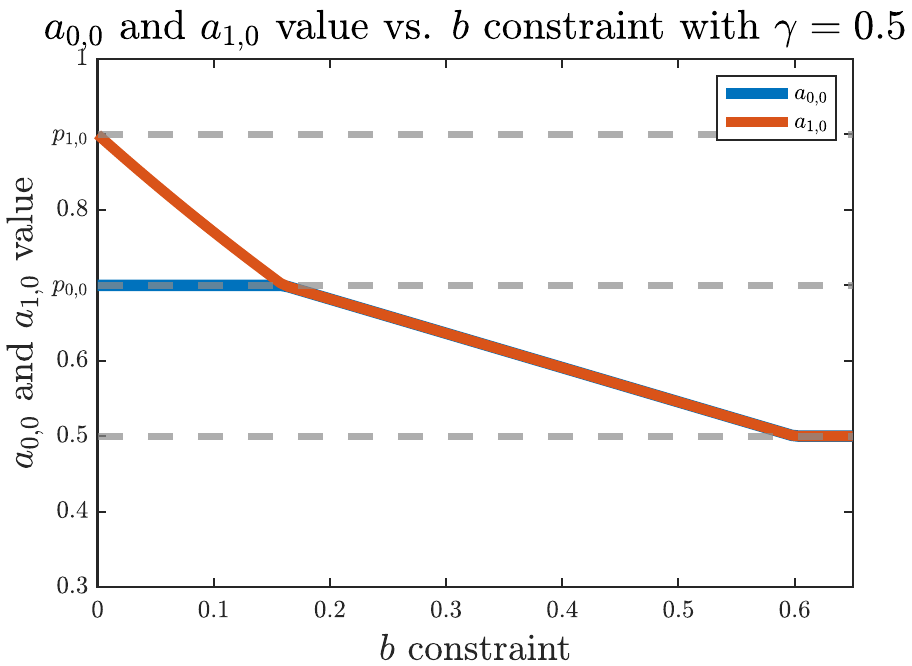}
        \caption{$p_{0,0}, p_{1,0} \geq \frac{1}{2}$}
        \label{fig:p0p1sameside}
    \end{subfigure}
    \begin{subfigure}[b]{0.32\textwidth}
        \centering
        \includegraphics[width=\textwidth]{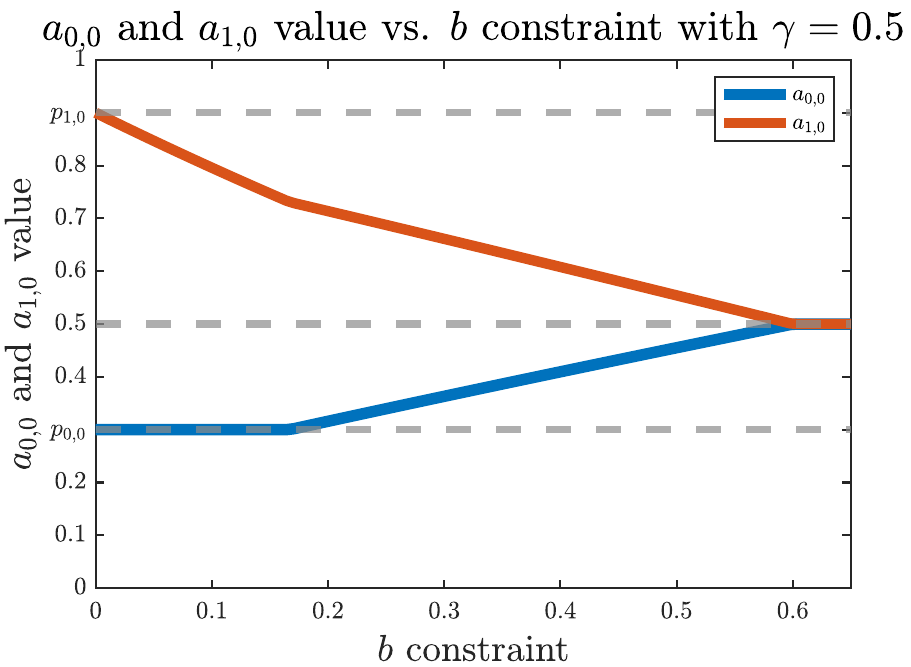}
        \caption{$p_{1,0} > \frac{1}{2}> p_{0,0} $}
        \label{fig:p0p1diffside}
    \end{subfigure}
    \begin{subfigure}[b]{0.32\textwidth}
        \centering
        \includegraphics[width=\textwidth]{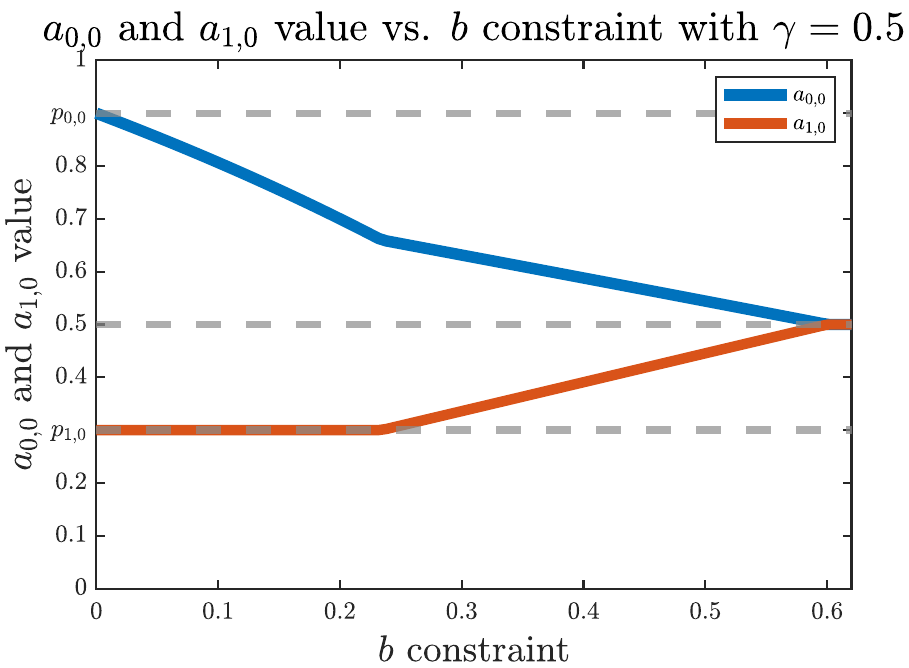}
        \caption{$p_{0,0} > \frac{1}{2}> p_{1,0} $}
        \label{fig:p0p1diffside_1}
    \end{subfigure}
    \caption{$a_{0,0}, a_{1,0}$ value vs. b constraint with $\mathbf{d} = [0.5,0.5]$ and $\gamma = 0.5$}
    \label{fig:plota0a1}
\end{figure}

In this section, we provide numerical results to illustrate the results given in Theorems \ref{thm:p0p1sameside} and \ref{thm:p0p1diffside}. In all panels in Figure \ref{fig:plota0a1}, the red solid curve represents the action in state $1$, denoted by $a_{1,0}$, while the blue curve represents the action in state $0$, denoted by $a_{0,0}$.

The behavior of the solution is, in a sense, similar to a water-filling strategy for Theorem \ref{thm:p0p1sameside} in Figure \ref{fig:p0p1sameside}. Since $p_{1,0}$ is farther from $\frac{1}{2}$ than $p_{0,0}$, in the strongly constrained setting (regime I) only state $1$ becomes active state. In other words, the optimal approach is to hold $\rva_0=\rvp_0$ and reduce $a_{1,0}$ from $p_{1,0}$ toward $\frac{1}{2}$. When $a_{1,0}$ decreases to $p_{0,0}$, i.e.,, for $b=b_1$ (regime II), the two states become identical. From that point forward, $\rva_0$ and $\rva_1$ must be tuned together, with $\rva_0=\rva_1$, until both attain $\frac{1}{2}$ (regime III). 

In Figure \ref{fig:p0p1diffside}, the difference in this case is that, in the second regime, the optimal policy $\rva_0$ and $\rva_1$ are not the same, as given in Theorem \ref{thm:p0p1diffside} case $p_{1,0} \geq \frac{1}{2} \geq p_{0,0}$. In this scenario, state $1$ becomes active first in regime I since $p_{1,0}$ lies farther from $\frac{1}{2}$ than state $0$. Once $b$ reaches $b = b_1^-$, both states become active state and gradually approach $\frac{1}{2}$ as $b$ increases. The final figure \ref{fig:p0p1diffside_1} illustrates the reverse situation, namely $p_{0,0} \geq \frac{1}{2} \geq p_{1,0}$, exhibiting the same trend as in figure \ref{fig:p0p1diffside}.

\section{Optimal Policy for the Binary Second-Order CMDP}

We now consider the case with $(n=2, k = 2)$, i.e., a language model with a vocabulary of size 2 that employs a sliding window of length 2. Now the  state space is $\mathsf{S} := \{(s^{(1)}, s^{(2)}) \mid s^{(i)} \in \{0,1\} \} = \{(0,0), (0,1), (1,0), (1,1)\}$, where $s^{(1)}$ is the current token and $s^{(2)}$ is the previous token. In this setup, $$\Gamma(\rvs) = \{(i,s^{(1)}):i \in \{ 0,1\}, s^{(1)} \text{ first element of }\rvs \},$$ and $\mathsf{A}(\rvs) = \{ \rva \in \Delta_4 : a_{\rvs'} = 0 \; \forall \rvs' \notin \Gamma(\rvs) \}$. An illustration of the $(2,2)$ setting is shown in Figure \ref{fig:(2,2)_illustrate}.

We characterize $\Gamma(\rvs)$ via two functions \(n_0(\cdot), n_1(\cdot): \mathsf{S} \rightarrow \mathsf{S}\), each of which maps the current state to the corresponding next state. Specifically, \(n_0\) is applied when the next token is \(0\), and \(n_1\) is applied when the next token is \(1\). The complete list of input–output pairs is:
\begin{align}
    \begin{matrix}
    n_0((0,0)) = (0,0)\\
    n_0((0,1)) = (0,0)\\
    n_0((1,0)) = (0,1)\\
    n_0((1,1)) = (0,1)
    \end{matrix}
    ,\quad
    \begin{matrix}
    n_1((0,0)) = (1,0)\\
    n_1((0,1)) = (1,0)\\
    n_1((1,0)) = (1,1)\\
    n_1((1,1)) = (1,1)
    \end{matrix}.
\end{align}

\begin{figure}
    \centering
    \includegraphics[width=0.5\linewidth]{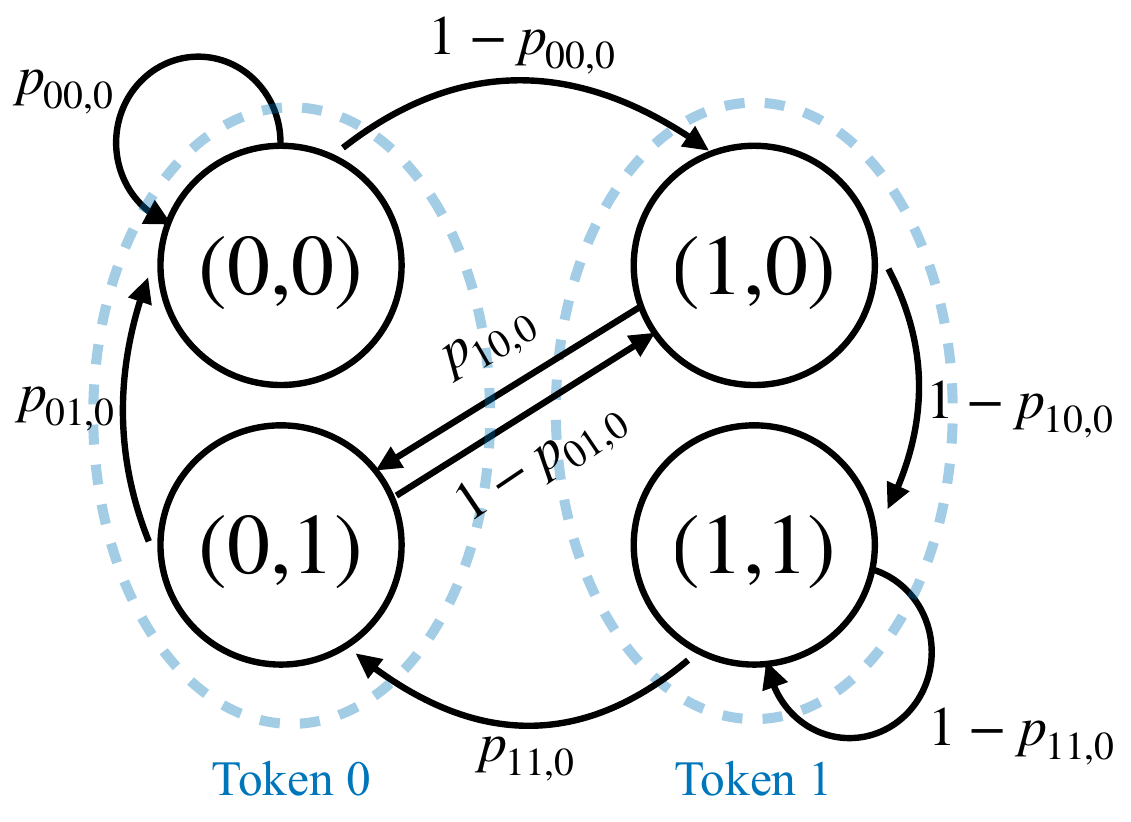}
    \caption{State transition model of $(n,k) = (2,2)$ case.} 
    \label{fig:(2,2)_illustrate}
\end{figure}

 Thus, even though the model has 4 states in total, only two transition probabilities from each state are non-zero. Recall the transition probability $p_{\rvs,i}$ introduced in \eqref{eq:def_p_s,i}. The transition matrix can then be expressed as
\begin{align}
    \mP =
    \begin{bmatrix}
        \rvp_{00}\\
        \rvp_{01}\\
        \rvp_{10}\\
        \rvp_{11}
    \end{bmatrix} =
    \begin{bmatrix}
        p_{00,0} & 0              & p_{00,1} & 0 \\
        p_{01,0} & 0              & p_{01,1} & 0 \\
        0             & p_{10,0} & 0             & p_{10,1} \\
        0             & p_{11,0} & 0             & p_{11,1}
    \end{bmatrix} =
    \begin{bmatrix}
        p_{00,0} & 0 & 1-p_{00,0} & 0 \\
        p_{01,0} & 0 & 1-p_{01,0} & 0 \\
        0 & p_{10,0} & 0 & 1-p_{10,0} \\
        0 & p_{11,0} & 0 & 1-p_{11,0}
    \end{bmatrix}.
\end{align}
Here, we simplify the subscript notation ${(s^{(1)},s^{(2)})}$ by writing it as $s^{(1)}s^{(2)}$; for instance, we write $\rvp_{(0,0)}$ as $\rvp_{00}$. Consequently, the transition matrix $\mP$ is fully specified by just four parameters, $p_{\rvs,0}$. Similarly, the action $\rva_\rvs$ in $\mathsf{A}(\rvs)$, represented by the action matrix $\mA$, also depends on four independent variables $a_{\rvs,0}$ and can be written as
\begin{align}
    \mA =
    \begin{bmatrix}
        \rva_{00}\\
        \rva_{01}\\
        \rva_{10}\\
        \rva_{11}
    \end{bmatrix} =
    \begin{bmatrix}
        a_{00,0} & 0 & 1-a_{00,0} & 0 \\
        a_{01,0} & 0 & 1-a_{01,0} & 0 \\
        0 & a_{10,0} & 0 & 1-a_{10,0} \\
        0 & a_{11,0} & 0 & 1-a_{11,0}
    \end{bmatrix}.
\end{align}
With this setup in place, we can express problem $\mathbf{P3}$ as the following problem, denoted by $\mathbf{P3\text{-}(2,2)}$:
\begin{align}
    \mathbf{P3\text{-}(2,2)}: ~ \max_{d_{\rvs,0}, x_{\rvs,0}}  & \; \; \sum_{\rvs \in \mathsf{S}} d_{\rvs,0} H\!\left(\tfrac{x_{\rvs,0}}{d_{\rvs,0}}\right) 
    \label{eq:x0_obj_func_TV_4s} \\
    \text{subject to:} & ~ d_{00,0} =\rvd_{00}^{\gamma}+ \gamma( x_{00,0} + x_{01,0}) \label{eq:x0_constraint1_TV_4s}\\
    & ~ d_{01,0} =\rvd_{01}^{\gamma}+ \gamma( x_{10,0} + x_{11,0}) \label{eq:x0_constraint2_TV_4s}\\
    & ~ d_{10,0} =\rvd_{10}^{\gamma}+ \gamma ( d_{00,0}- x_{00,0}) + \gamma ( d_{01,0}- x_{01,0})
    \label{eq:x0_constraint3_TV_4s} \\
    & ~ d_{11,0} =\rvd_{11}^{\gamma}+ \gamma ( d_{10,0}- x_{10,0}) + \gamma ( d_{11,0}- x_{11,0})
    \label{eq:x0_constraint4_TV_4s} \\
    &  ~ \sum_{\rvs \in \mathsf{S}} \big| d_{\rvs,0} p_{\rvs,0}- x_{\rvs,0} \big|\leq \frac{b}{2}\label{eq:x0_constraint5_TV_4s}\\
    & ~ 0 \leq x_{\rvs,0} \leq d_{\rvs,0}, \; \forall \rvs \in \mathsf{S}\label{eq:x0_constraint6_TV_4s}
\end{align}
where we define $a_{\rvs,0} = \tfrac{x_{\rvs,0}}{d_{\rvs,0}}$. Because the optimal action $a_{\rvs,0}$ varies with the value of $b$, we may express it explicitly as a function of $b$, denoted $a_{\rvs,0}(b)$, and define the corresponding action matrix as $\mA(b)$.

\subsection{Motivating Numerical Examples of $\mathbf{P3\text{-}(2,2)}$} \label{sec:numerical_example_4s}

We begin by presenting several representative numerical solutions solved by CVX package to show how the optimal policy changes as the TV budget \(b\) increases. These examples serve to motivate the regime classification; a more detailed analysis appears in section \ref{sec:numerical_example_4s_1}.

The numerical experiment shown in Figure \ref{fig:P3_22_six_fig} closely resembles the behavior observed in the binary first-order state case $(n,k) = (2,1)$. We observe that in all scenarios, when $b$ is zero, the actions coincide with the original transition probabilities. As $b$ increases, initially only one state's action $a_{\rvs,0}$ starts to move apart from $p_{\rvs,0}$, and the number of active states grows with larger values of $b$. Eventually, all four states have active actions $a_{\rvs,0}$, and the system reaches an unconstrained regime in which the actions exhibit a uniform transition probability. Accordingly, we split the solution regimes based on the number of active states. 

\subsection{Solution Regimes for $\mathbf{P3\text{-}(2,2)}$}

The solution to the $\mathbf{P3\text{-}(2,2)}$ problem can be partitioned into five regimes based on the budget level $b$. The regime index $\mathrm{I - IV}$ corresponds to the number of active states, except in regime V, which is an unconstrained case whose solution is the uniform distribution over the feasible next states.

\begin{enumerate}
    \item Regime I: Highly constrained regime \\
    Exactly one active state. 
    \item Regime II–IV: Intermediately constrained regimes \\
    In these regimes, II, III, and IV exhibit two, three, and four active states, respectively.
    \item Regime V: Unconstrained regime \\
    For all four states, the transition distribution is uniform over the possible subsequent states.
\end{enumerate}

The regime transition conditions are summarized in Table \ref{tab:four_state_regime_boundaries}; $\rvs_i$ denotes the $i$th state to activate, and $\mathcal{M}$ denotes the set of active states. The boundary conditions are expressed as equations that indicate precisely when the next additional state becomes active. We provide precise definitions of these variables in the table below:
\begin{align}
    & I_{\rvs}^{\lambda}(\mA)  = |K^\lambda_{\rvs }(\mA)|, 
    \quad K^{\lambda}_{\rvs}(\mA)  =  \log \frac{a_{\rvs,0}}{1-a_{\rvs,0}}- \gamma \big([\mW^\lambda(\mA)]_{n_0(\rvs)}- [\mW^\lambda(\mA)]_{n_1(\rvs)}\big) \\
    & \text{where } \mW^{\lambda}(\mA) = \mM(b)^\top \big[ \mH(\mA)-\lambda \mD(\mA,\mP) \big], ~ \mM(b)  =  \left(I-\gamma \mA(b)^{\top}\right)^{-1} \notag \\
    & \text{and }\mD(\mA,\mP)  = \begin{bmatrix}
        |a_{00,0} - p_{00,0}| \\
        |a_{01,0} - p_{01,0}| \\
        |a_{10,0} - p_{10,0}| \\
        |a_{11,0} - p_{11,0}|
    \end{bmatrix}, ~ \mH(\mA) := 
    \begin{bmatrix}
        H(\rva_{00}) \\ H(\rva_{01}) \\ H(\rva_{10}) \\ H(\rva_{11})  
    \end{bmatrix} =
    \begin{bmatrix}
        H(a_{00,0}) \\ H(a_{01,0}) \\ H(a_{10,0}) \\ H(a_{11,0})  
    \end{bmatrix}.
\end{align}
Here, $H(a_{\rvs,0})$ represents the binary entropy function, because each state-transition probability vector contains exactly two non-zero entries. $\lambda$ denotes the dual variable corresponding to the unnormalized total-variation budget constraint.

\begin{table}[]
    \centering
    \caption{Four-state CMDP: regime constraints and boundary conditions.}
    \label{tab:four_state_regime_boundaries}
    \small
    \setlength{\tabcolsep}{2pt}
    \renewcommand{\arraystretch}{1.25}
    \begin{tabular}{@{}
        p{0.17\columnwidth}
        p{0.22\columnwidth}
        p{0.55\columnwidth}
        @{}}
        \hline
        Regime
        &
        Active state set $\mathcal{M}$
        &
        Regime constraints and terminal boundary
        \\
        \hline

        \(\mathrm{I}\),
        \(b\in[0,b_1)\)
        &
        \(\mathcal M_{\mathrm I}
        =\{\rvs_1\}\)
        &
        Regime constraints: \par
        \(
        \begin{aligned}
             I_{\rvs}^{\lambda(b)}(\mA(b))
            <I_{\rvs_1}^{\lambda(b)}(\mA(b)),
            \forall \rvs\neq\rvs_1.
        \end{aligned}
        \)
        \par\smallskip
        Boundary \(b_1\) happens when :\par
        \(
        \begin{aligned}
            I_{\rvs_2}^{\lambda(b_1)}(\mA(b_1))
            &=
            I_{\rvs_1}^{\lambda(b_1)}(\mA(b_1))\\
            I_{\rvs}^{\lambda(b_1)}(\mA(b_1))
            &<I_{\rvs_1}^{\lambda(b)}(\mA(b_1)), \rvs\notin\{\rvs_1,\rvs_2\}.
        \end{aligned}
        \)
        \\
        \hline

        \(\mathrm{II}\),
        \(b\in[b_1,b_2)\)
        &
        \(\mathcal M_{\mathrm{II}}
        =\{\rvs_1,\rvs_2\}\)
        &
        Regime constraints: \par
        \(
        \begin{aligned}
            & I_{\rvs_1}^{\lambda(b)}(\mA(b))
            =
            I_{\rvs_2}^{\lambda(b)}(\mA(b)), \\ 
            & I_{\rvs}^{\lambda(b)}(\mA(b))
            <I_{\rvs_1}^{\lambda(b)}(\mA(b)),
            ~ \rvs\notin\{\rvs_1,\rvs_2\}.
        \end{aligned}
        \)
        \par\smallskip
        Boundary \(b_2\) happens when: \par
        \(
        \begin{aligned}
            I_{\rvs_1}^{\lambda(b_2)}(\mA(b_2))
            &=
            I_{\rvs_2}^{\lambda(b_2)}(\mA(b_2)) =
            I_{\rvs_3}^{\lambda(b_2)}(\mA(b_2)) \\
            I_{\rvs}^{\lambda(b_2)}(\mA(b_2))
            &<I_{\rvs_1}^{\lambda(b_2)}(\mA(b_2)), ~ \rvs \notin \{ \rvs_1, \rvs_2, \rvs_3\}.
        \end{aligned}
        \)
        \\
        \hline

        \(\mathrm{III}\),
        \(b\in[b_2,b_3)\)
        &
        \(\mathcal M_{\mathrm{III}}
        =\{\rvs_1,\rvs_2,\rvs_3\}\)
        &
        \text{Regime constraints:}\par
        \(
        \begin{aligned}
            I_{\rvs_1}^{\lambda(b)}(\mA(b))
            &=
            I_{\rvs_2}^{\lambda(b)}(\mA(b))=
            I_{\rvs_3}^{\lambda(b)}(\mA(b))\\
            I_{\rvs}^{\lambda(b)}(\mA(b))
            &<I_{\rvs_1}^{\lambda(b)}(\mA(b)), ~ \rvs \notin \mathcal{M}_\mathrm{III}.
        \end{aligned}
        \)
        \par\smallskip
        \text{Boundary \(b_3\) happens when:}\par
        \(
        \begin{aligned}
            I_{\rvs}^{\lambda(b_3)}(\mA(b_3))=
            I_{\rvs'}^{\lambda(b_3)}(\mA(b_3)), \forall \rvs, \rvs' \in \mathsf{S}.
        \end{aligned}
        \)
        \\
        \hline

        \(\mathrm{IV}\),
        \(b\in[b_3,b_4)\)
        &
        \(\mathcal M_{\mathrm{IV}}
        =\mathsf S\)
        &
        \text{Regime constraints:}\par
        \(
        I_{\rvs}^{\lambda(b)}(\mA(b)) >0,
        \quad
        \forall\rvs\in\mathsf S.
        \)
        \par\smallskip
        \textit{Boundary \(b_4\) happens when:}\par
        \(
        \begin{aligned}
            I_{\rvs}^{\lambda(b_4)}(\mA(b_4))=0,
            \quad \forall\rvs\in\mathsf S.
        \end{aligned}
        \)
        \\
        \hline
    \end{tabular}
\end{table}

\subsection{Regime I Solution to $\mathbf{P3\text{-}(2,2)}$}

\begin{figure*}[t]
    \centering

    \begin{subfigure}[t]{0.31\textwidth}
        \vspace{0pt}
        \centering
        \includegraphics[width=\textwidth]
        {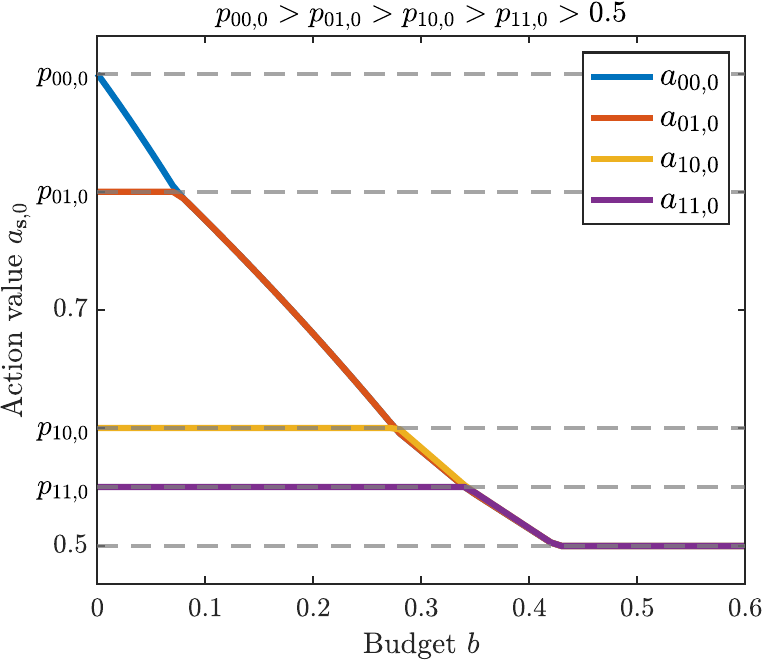}
        \subcaption{$p_{\rvs,0} \geq 0.5$ for all four states.}
        \label{fig:case_a}
    \end{subfigure}
    \hfill
    \begin{subfigure}[t]{0.31\textwidth}
        \vspace{0pt}
        \centering
        \includegraphics[width=\textwidth]
        {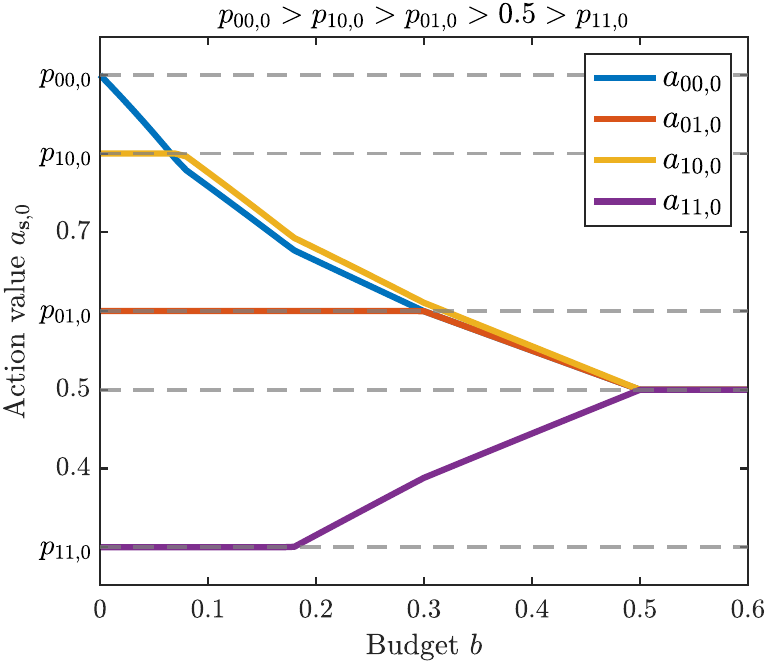}
        \subcaption{One value of $p_{\rvs,0}$ lies on the opposite side of $0.5$.}
        \label{fig:case_d}
    \end{subfigure}
    \hfill
    \begin{subfigure}[t]{0.31\textwidth}
        \vspace{0pt}
        \centering
        \includegraphics[width=\textwidth]
        {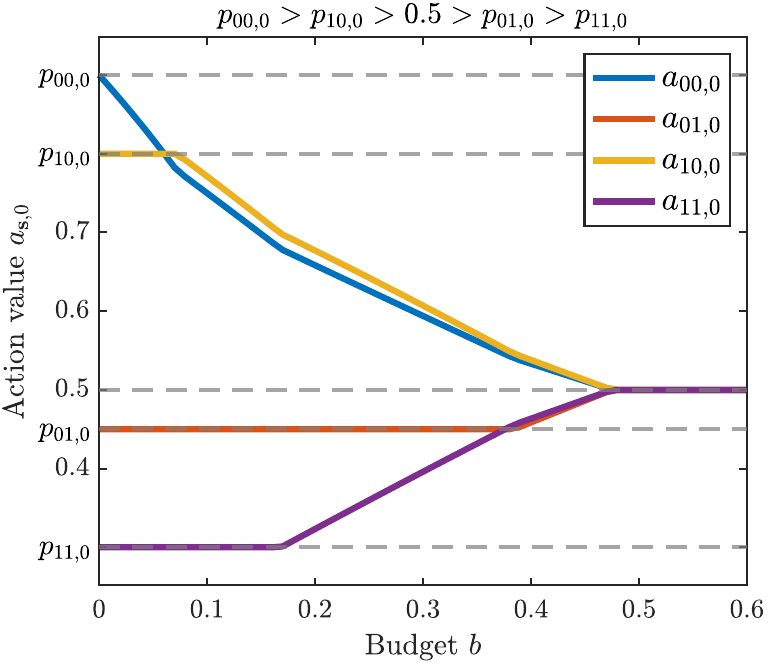}
        \subcaption{Two values of $p_{\rvs,0}$ lie on the opposite side of $0.5$.}
        \label{fig:case_e}
    \end{subfigure}
    \caption{Optimal policies for problem
    $\mathbf{P3\text{-}(2,2)}$ with $\gamma=0.5$ and
    $\rvd=[0.25,0.25,0.25,0.25]$.}
    \label{fig:P3_22_six_fig}
\end{figure*}

For $b = 0$, it follows that the optimal action matrix $\mA(b) = \mP$, and hence $\mD(\mA,\mP) = \mathbf{0}$. Thus, when $b = 0$, we rename the variables as follows:
\begin{align}   
    I^0_{\rvs}(\mP) &:= \big| K^0_{\rvs}(\mP)\big|, \qquad 
    K^0_{\rvs}(\mP) = \log\frac{p_{\rvs, 0}}{1-p_{\rvs, 0}}
    - \gamma\Big([\mW^0(\mP)]_{n_0(\rvs)} - [\mW^0(\mP)]_{n_1(\rvs)}\Big) \label{eq:I^0_def_P3_22}\\
    \mW^0(\mP) & := \mM(0)^\top \mH(\mP) = (I-\gamma \mP)^{-1} \mH(\mP) 
\end{align}

\begin{theorem}[$\mathbf{P3\text{-}(2,2)}$ Optimal policy for regime I]\label{thm:4s_regimeI}
    Let $\rvs_1 = \argmax_{\rvs \in \mathsf{S}} I^0_{\rvs}(\mP)$ and define  $\sigma_{\rvs} = \sign(p_{\rvs,0}-0.5)$. Suppose that this $\rvs_1$ is unique and $\sigma_{\rvs_1} K^0_{\rvs_1}(\mP) > 0$, 
    then when $b \in [0,b_1)$, the optimal policy is given by
    \begin{equation}
            a_{\rvs,0}
            =
            \begin{cases}
                p_{\rvs,0}- \sigma_{\rvs} u_{\rvs_1}(b),
                &
                \rvs=\rvs_1,
                \\
                p_{\rvs,0},
                &
                \text{otherwise},
            \end{cases}
    \end{equation}
    where
    \begin{align}
        u_{\rvs_1}(b) & = \frac{b}{2\rvd^{(base)}_{\rvs_1}+ b \gamma\sigma_{\rvs_1} \Delta \mM(0)_{\rvs_1}}, \quad \rvd^{(base)} = \mM(0)\,\rvd^{\gamma}, \\
        \Delta \mM(0)_{\rvs_1}
        & = \left[\mM(0)\right]_{\rvs_1,n_1(\rvs_1)}
        -\left[\rmM(0)\right]_{\rvs_1,n_0(\rvs_1)}.
    \end{align}
\end{theorem}
In Theorem \ref{thm:4s_regimeI}, $b_1$ denotes the smallest value of $b$ at which the boundary condition in Table \ref{tab:four_state_regime_boundaries} is met. More specifically, we define
\begin{align}
    b_1 = \inf\{b>0 : I^{\lambda(b)}_{\rvs_1}(\mA(b)) = \max_{\rvs \neq \rvs_1} I^{\lambda(b)}_{\rvs}(\mA(b))\}, \label{eq:b1_def_thm5_4s}
\end{align}
where $\lambda(b)$ denotes the dual variable associated with the constraint in \eqref{eq:x0_constraint5_TV_4s}, and its value depends on the particular choice of $b$. The expression for $\lambda(b)$ is given in Appendix C,
see equation (230).

This theorem provides the closed-form expression of the optimal policy for regime I, where only one state is active. The solution is derived by finding the root of the unnormalized total variation budget constraint, with the full proof presented in Appendix C. 
The other regimes are more complex, and a closed-form solution has not been found for them.

\subsection{Numerical Experiments for Binary Second-Order CMDP}\label{sec:numerical_example_4s_1}

\begin{figure}[t]
    \centering

    \begin{subfigure}[t]{0.48\columnwidth}
        \centering
        \includegraphics[width=\linewidth]{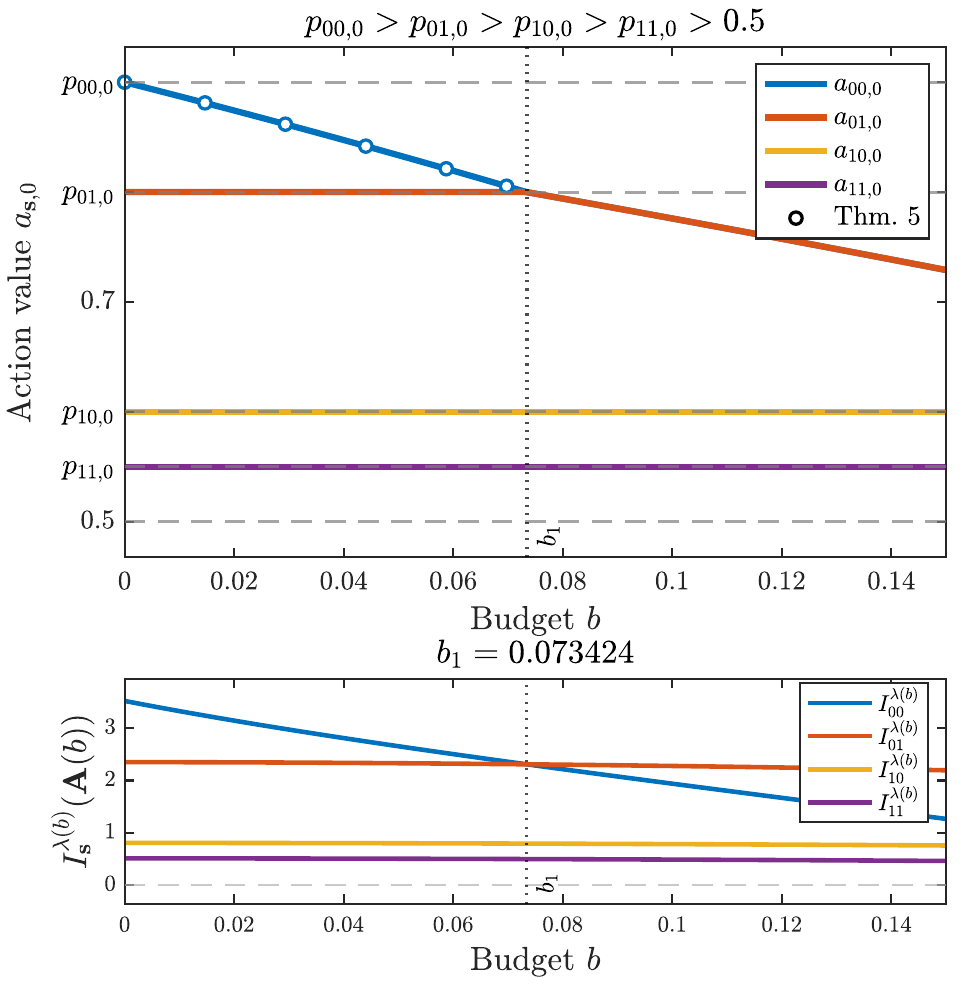}
        \caption{Zoomed-in view of Figure \ref{fig:case_a}.}
        \label{fig:thm_4s_regimeIa}
    \end{subfigure}
    \hfill
    \begin{subfigure}[t]{0.48\columnwidth}
        \centering
        \includegraphics[width=\linewidth]{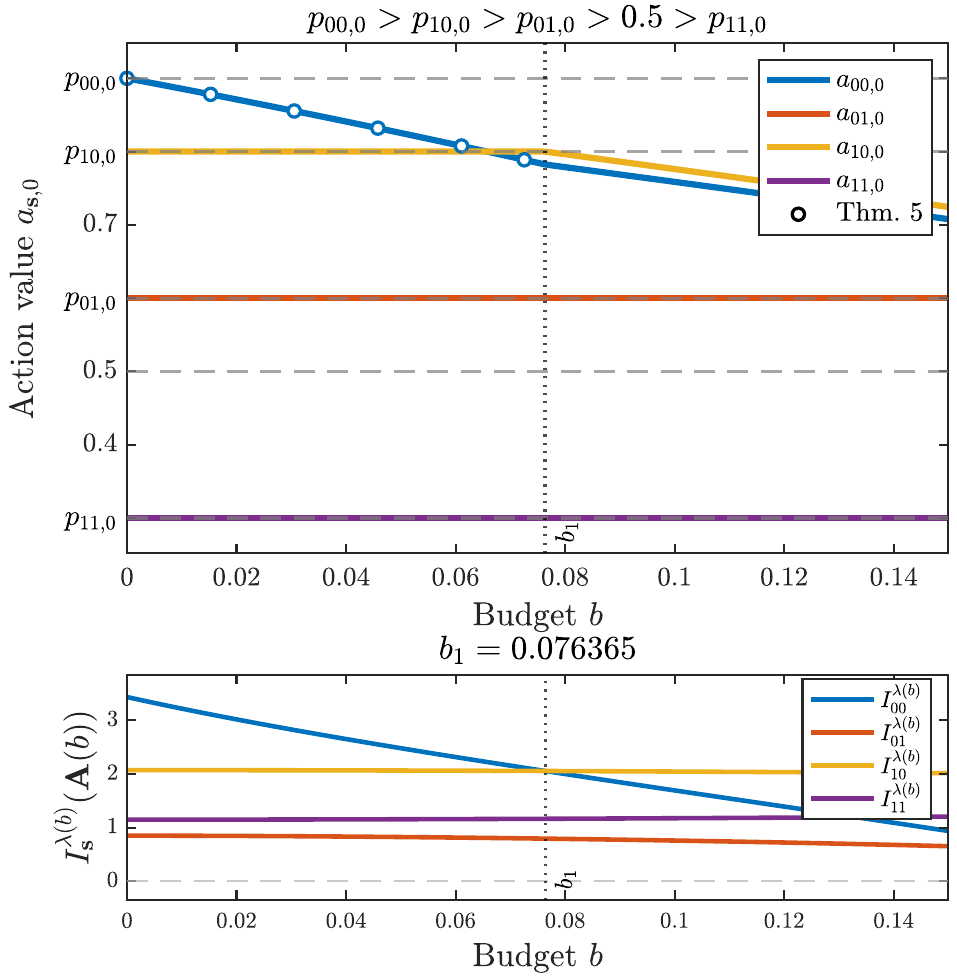}
        \caption{Zoomed-in view of Figure \ref{fig:case_d}.}
        \label{fig:thm_4s_regimeIb}
    \end{subfigure}

    \caption{Problem $\mathbf{P3\text{-}(2,2)}$ Regime-I optimal policy and its $b_1$ boundary.}
    \label{fig:thm_4s_regimeI_example}
\end{figure}

Figure \ref{fig:thm_4s_regimeI_example} presents two examples that illustrate the boundary $b_1$ and the closed-form solution for regime I. These examples revisit the numerical experiments illustrated in Figures \ref{fig:case_a} and \ref{fig:case_d}. In the upper panel, the circular marker indicates the closed-form solution given by Theorem \ref{thm:4s_regimeI} in regime I, whereas the solid line shows the optimal solution obtained using the CVX package. Moreover, the lower panel displays the values of $I^{\lambda(b)}_{\rvs}$ for different states $\rvs$. According to equation \eqref{eq:b1_def_thm5_4s}, the boundary of regime I is identified at the smallest value of $b$ for which the maximum of $I^{\lambda(b)}_{\rvs}$ becomes equal to $I^{\lambda(b)}_{\rvs_1}$.

In both figures, the first active state is $\rvs_1 = (0,0)$. As $b$ increases and reaches the threshold $b_1$, the point in the lower panel where the blue value $I^{\lambda(b)}_{00}$ intersects the other $I^{\lambda(b)}_{\rvs}$ marks the activation of a second state. In Figure \ref{fig:thm_4s_regimeIa}, this second active state is $(0,1)$ (shown in red), and the action $a_{01,0}$ begins to deviate from $p_{01,0}$ once $b$ exceeds $b_1$. A similar pattern appears in Figure \ref{fig:thm_4s_regimeIb}, where the second active state is $(1,0)$ (shown in yellow), and its corresponding action $a_{10,0}$ likewise starts to diverge from $p_{10,0}$ after $b_1$.

\section{Numerical Studies for the Ternary First-Order CMDP}

For the setting $(n = 3, k = 1)$, the state space reduces to $\mathsf{S} = \{0,1,2\}$, and $\Gamma(s) = \mathsf{S}$ holds for all $s \in \mathsf{S}$. We write $s$ rather than $\rvs$ here since each state corresponds only to the current token, rather than a vector formed by stacking the previous $k$ tokens. Consequently, $\mathsf{A}(s) = \Delta_3$. Substituting these sets into the optimization problem $\mathbf{P3}$, we can write it as $\mathbf{P3\text{-}(3,1)}$ below:
\begin{align}
    \mathbf{P3\text{-}(3,1)}: ~ \max_{\substack{(\rvx_s , d_s)_{s \in \{0,1,2 \} } \\
        \rvx_s = [x_{s, s'}]_{s' \in \{0,1,2 \}}}}  & \; \; d_0 H\!\left(\tfrac{\rvx_0}{d_0}\right) + d_1 H\!\left(\tfrac{\rvx_1}{d_1}\right) + d_2 H\!\left(\tfrac{\rvx_2}{d_2}\right) \label{eq:x0_obj_func_TV_3s} \\
    \text{subject to:} & ~  d_s = \rvd^\gamma_s + \gamma (x_{0,s} + x_{1,s} + x_{2,s}), \forall s \in \mathsf{S} \label{eq:x0_constraint_1-0_TV_3s}\\
    & ~ \sum_{s=0}^{2} d_s \overline{\TV}_s (\tfrac{\rvx_s}{d_s}) \leq b \label{eq:x0_constraint_2_TV_3s}\\
    & ~ 0 \leq x_{s,i} \leq d_s , \text{ and } \sum_{i=0}^2 x_{s,i} = d_s , \; \forall s,i \in \{ 0,1,2\} \label{eq:x0_constraint_3_TV_3s}
\end{align}

The three-state case is significantly more complex than both binary CMDP cases. Solving this $\mathbf{P3\text{-}(3,1)}$ problem requires addressing at most $6$ independent actions, which exceeds the $2$ independent actions in the $(2,1)$ case and the $4$ independent actions in the $(2,2)$ case.
When adjusting the state transition probabilities $\rvp_s$ to $\rva_s$, the situation falls into two scenarios: either two transition probabilities for that state change, or all three change.  
To make this explicit, we introduce the following definition. 
\begin{definition}[Pair-active state, fully-active state] \label{def:active-state-(3,3)}
    For the case $(n=3, k=1)$, under a given unnormalized total variation budget $b$, a state is called a pair-active state if the difference $\rve_s := \rva_s - \rvp_s$ contains exactly one zero entry. 
    The state is called a fully-active state if the difference $\rve_s$ has no zero entry.
\end{definition}
For instance, the adjustment in one active state may take the form
\begin{equation}
    \rve_s = 
    (u_s,-u_s,0), \text{ where } u_s \neq 0, \label{eq:pair-active-def}
\end{equation}
or
\begin{equation}
    \rve_s = 
    (u_s,v_s,-u_s-v_s), \text{ where } u_s, v_s  \neq 0 \text{ and } u_s + v_s \neq 0.  \label{eq:fully-active-def}
\end{equation}
We refer to \eqref{eq:pair-active-def} as a pair-active state, and to \eqref{eq:fully-active-def} as a fully-active state.

In contrast to the binary cases, the ternary model permits several distinct configurations of pair-active and fully-active states, leading to more complex optimal policy behaviors. Accordingly, instead of attempting a full analytical description of every possible active-state configuration, we concentrate on numerical examples to demonstrate these behaviors.

\subsection{Numerical Examples of $\mathbf{P3\text{-}(3,1)}$} \label{sec:numerical_example_3s}

\begin{figure}[t]
    \centering
    \begin{subfigure}[t]{0.47\columnwidth}
        \centering
        \includegraphics[width=\linewidth]
        {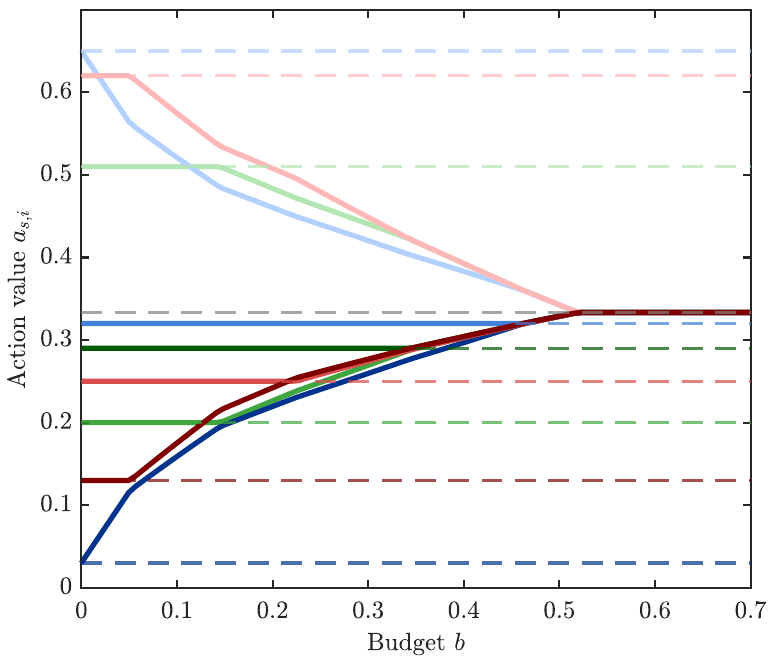}
        \subcaption{Transition probability: $\rvp_0 =[0.51, 0.2, 0.29], \\ \rvp_1 =[0.65, 0.32, 0.03], \rvp_2 =[0.62, 0.25, 0.13]$.}
        \label{fig:case_a_3s}
    \end{subfigure}
    \hfill
    \begin{subfigure}[t]{0.47\columnwidth}
        \centering
        \includegraphics[width=\linewidth]
        {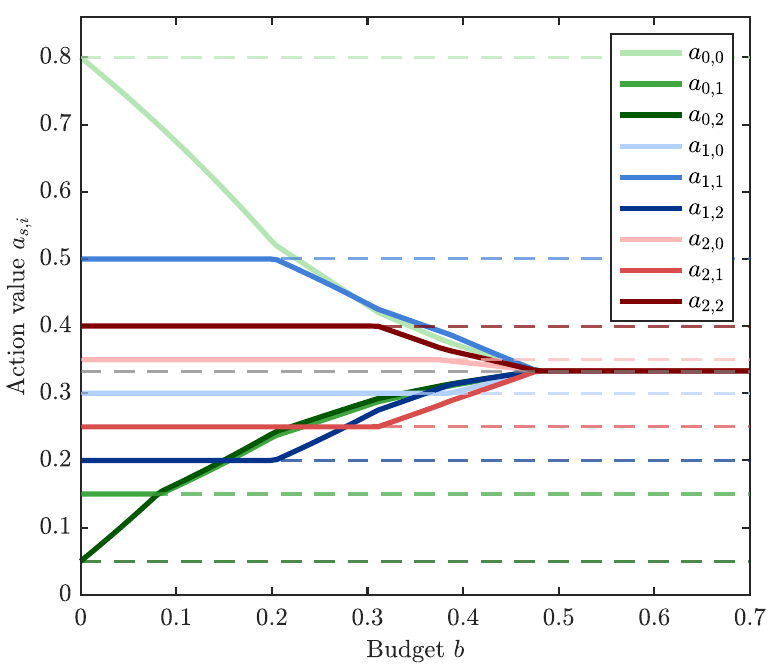}
        \subcaption{Transition probability: $\rvp_0 =[0.8, 0.15, 0.05], \\ \rvp_1 =[0.3, 0.5, 0.2], \rvp_2 =[0.35, 0.25, 0.4]$.}
        \label{fig:case_d_3s}
    \end{subfigure}
    \caption{Optimal policies for problem
    $\mathbf{P3\text{-}(3,1)}$ with $\gamma=0.5$ and
    $\rvd=[\tfrac{1}{3}, \tfrac{1}{3}, \tfrac{1}{3}]$.}
    \label{fig:P3_31_six_fig}
\end{figure}

We provide two numerical examples, and the solutions are computed using the CVX numerical optimization solver to illustrate how the optimal policy evolves as the TV budget \(b\) increases. 
The colored dashed lines in Figure \ref{fig:P3_31_six_fig} represent the original transition probabilities $p_{s,i}$ from state $s$ to state $i$, while the colored solid lines indicate the corresponding optimal actions. The gray dashed line denotes the uniform distribution level ($\frac{1}{3}$) as a reference. Green is used for state $0$, blue for state $1$, and red for state $2$.

It can seen that as \(b\) grows from $0$, a state first becomes pair-active; for example, in Figure \ref{fig:case_a_3s}, for state \(1\), the actions \(a_{1,0}\) and \(a_{1,2}\), shown in light and dark blue, are the first to become active when $b$ is small, whereas \(a_{1,1}\) remains unchanged. As $b$ further increases, two types of transitions can occur: a pair-active state turns into a fully active state, or a non-active state turns into a pair-active state. In figure \ref{fig:case_a_3s}, the blue state (state $1$) becomes pair-active first, followed by the red state (state $2$) becoming pair-active as $b$ increases; in contrast, in figure \ref{fig:case_d_3s}, the green state (state $0$) initially becomes pair-active and then transitions to fully-active state. These examples illustrate the more complex behavior when the number of states increases from two to three.

\section{Conclusion}
In this study, we presented a novel formulation for relatively-secure LLM-based steganography in the Constrained Markov Decision Process (CMDP) framework. This approach considers the impact on both the immediate and future token generations, and led us to insights into how the optimal policy should be chosen. The solution suggests that when choosing the adjustment of the probability distribution, the choice should be made considering the LLM-generated transition probabilities at all states, and usually adjusting it for the state that has the lower entropy should be prioritized. 

In this work, we abstract and simplify LLMs into a basic $(n,k)$ finite-state model, where $n$ denotes the vocabulary size and $k$ is the length of the preceding token context.We explicitly treat the cases $(2,1)$ and regime I of $(2,2)$. We also analyze numerical examples for the case $(3,1)$. In real-world settings, LLMs have much larger vocabularies $n$ and longer context lengths $k$, so the number of possible states $N = n^k$ grows exponentially. As a result, finding solutions becomes substantially more difficult, and many additional regimes must be taken into account. Nevertheless, this work provides an glimpse into how optimal policies may behave and evolve in these more complex models.

\section*{AI Use Disclosure}

ChatGPT (OpenAI) was used as a supporting tool in preparing this manuscript. It assisted in checking that notation was consistent across the paper. For the binary first-order \((2,1)\) CMDP analysis, it supported mathematical consistency checks, including the singleton-set argument and KKT conditions. For the binary second-order \((2,2)\) CMDP analysis, it aided with parts of the derivations and proof development. It was also used to improve grammar, clarity, and overall English. All AI-assisted mathematical material was independently reviewed and validated by the authors, who assume full responsibility for the results, arguments, and final presentation.

\bibliographystyle{IEEEtran}
\bibliography{stegaCMDP}

\newpage
\appendix
\input{appendix_TIT}

\end{document}

%% file: appendix_TIT.tex
\section{Proof of Theorem 1} \label{pf:thm1}
Assuming a solution $\Tilde{d}(\rvs,\cdot)$ exists for problem $\mathbf{P1}$, we can then construct a new occupancy measure $d^{\star}(\rvs,\cdot) =d_{\rvs}^{\star}\delta_{\rva_{\rvs}^{\star}}(\cdot)$, such that
\begin{align}
    d_{\rvs}^{\star}
    &=
    \int_{\mathsf{A}(\rvs)}
    \widetilde{d}(\rvs,\mathrm d\rva),
    \label{eq:d_star_measure}
    \\
    \rva_{\rvs}^{\star}
    &=
    \frac{1}{d_{\rvs}^{\star}}
    \int_{\mathsf{A}(\rvs)}
    \rva\,\widetilde{d}(\rvs,\mathrm d\rva)
    =
    [a_{\rvs,i}^{\star}]_{i\in\mathsf S},\text{ where }
    a_{\rvs,i}^{\star}
    =
    \frac{1}{d_{\rvs}^{\star}}
    \int_{\mathsf{A}(\rvs)}
    a_i\,\widetilde{d}(\rvs,\mathrm d\rva),
    \quad i\in \{0, 1, \dots, N-1 \} .
    \label{eq:a_star_measure}
\end{align}

where $a_i$ denotes the $i$-th entry of $\rva$ and $\delta_{\rva_{\rvs}^{\star}}(\cdot)$ denotes the Dirac measure
concentrated at $\rva_{\rvs}^{\star}$.
For any measurable set
$\mathcal B\subseteq \mathsf A(\rvs)$, the definition of $\delta_{\rva_{\rvs}^{\star}}(\mathcal B)$ is
\begin{equation}
    \delta_{\rva_{\rvs}^{\star}}(\mathcal B)
    =
    \begin{cases}
        1, & \rva_{\rvs}^{\star}\in\mathcal B,\\
        0, & \rva_{\rvs}^{\star}\notin\mathcal B.
    \end{cases}
\end{equation}

If $d_{\rvs}^{\star}=0$, $\rva_{\rvs}^{\star}$ may be chosen
arbitrarily from $\mathsf{A}(\rvs)$, since the corresponding
occupancy measure at state $\rvs$
has zero total mass.

We first verify that, under this definition, $\rva^{\star }_{s}$ remains in $\mathsf{A}(\rvs)$. To do this, it suffices to show that $a^{\star }_{s, i} = 0$ for all $i \notin \Gamma(\rvs)$. This follows from the fact that the integral defining $a^{\star }_{s, i}$ is taken only over the set where $\rva \in \mathsf{A}(\rvs)$. Consequently, for any such $i \notin \Gamma(\rvs)$ we have $a_i = 0$, and therefore $a^{\star }_{s, i} = 0$ as well.

We then need to verify that the occupancy measure $d^{\star} (\rvs,\cdot)$ meets the criteria outlined in constraints (\ref{eq:con_same1_TV}) to (\ref{eq:opt_problem_con4_TV}), and whether it results in a greater objective value for (\ref{eq:opt_problem_ob_TV}). 
\begin{enumerate}
    \item Constraint (\ref{eq:con_same1_TV}),(\ref{eq:opt_problem_con4_TV}): 
        $\tilde{d}(\rvs,\cdot)$ satisfies all constraints in $\mathbf{P1}$. In particular, for constraint (\ref{eq:opt_problem_con4_TV}), it is nonnegative for every $\tilde{d}(\rvs,\cdot)$. Consequently, each $d_\rvs^\star$ must be nonnegative for all $\rvs \in \mathsf{S}$, and the same holds for $d^\star(\rvs,\cdot)$. 

Thus, $d^\star(\rvs,\cdot)$ complies with constraints (\ref{eq:opt_problem_con4_TV}).
        Moreover, for each $\rvs \in \mathsf{S}$, 
        \begin{align}
            \int_{\mathsf{A(\rvs)}} d^\star(\rvs,\mathrm{d}\rva) & =  \int_{\mathsf{A(\rvs)}} d^\star_\rvs \delta_{\rva^\star_s}(\mathrm{d}\rva) = d^\star_\rvs = \int_{\mathsf{A(\rvs)}} \tilde{d} (\rvs,\mathrm{d}\rva) \\
            & = \rvd^\gamma_\rvs + \gamma \sum_{\rvs' \in \mathsf{S}} \int_{\mathsf{A(\rvs')}} a_{\rvs} \tilde{d}(\rvs',\mathrm{d}\rva) \label{eq:reduce_linear_con_pf_end} \\
            & = \rvd^\gamma_\rvs + \gamma \sum_{\rvs'\in \mathsf{S}} a^\star_{\rvs',\rvs} \;  d^\star_{\rvs'} = \rvd^\gamma_\rvs + \gamma \sum_{\rvs'\in \mathsf{S}} \int_{\mathsf{A(\rvs')}} a_{\rvs} \;  d^\star_{\rvs'}\delta _{\rva^\star_{\rvs'}}( \mathrm{d} \rva) \\
            & = \rvd^\gamma_\rvs + \gamma \sum_{\rvs'\in \mathsf{S}} \int_{\mathsf{A(\rvs')}} a_{\rvs} \;  d^\star(\rvs',\mathrm{d} \rva)
        \end{align}
        where the equality in (\ref{eq:reduce_linear_con_pf_end}) follows from the fact that $\Tilde{d} (\rvs,\cdot)$ satisfies the constraint (\ref{eq:con_same1_TV}). 
    \item Constraint (\ref{eq:opt_problem_con3_TV}):  For constraint (\ref{eq:opt_problem_con3_TV}), when $d_s^\star=0$, $\widetilde d(s,\cdot)$ becomes a zero measure, and states with zero mass contribute nothing to this constraint. Hence, we focus on the case $d_s^\star>0$ and obtain
    \begin{align}
        b & \geq \sum_{\rvs \in \mathsf{S}} \int_{\mathsf{A}(\rvs)} \overline{\TV}_{\rvs}(\rva)\tilde{d}(\rvs, \mathrm{d}\rva) \\
        & = \sum_{\rvs \in \mathsf{S}} d^\star_\rvs \int_{\mathsf{A}(\rvs)}   \overline{\TV}_{\rvs}(\rva) \left( \frac{\tilde{d} (\rvs,\mathrm{d}\rva)}{d^{\star }_{\rvs}} \right)
          \\
        & \geq \sum_{\rvs \in \mathsf{S}} d^{\star}_{\rvs} \overline{\TV}_{\rvs} \int_{\mathsf{A}(\rvs)} \rva \left( \frac{\tilde{d} (\rvs, \mathrm{d}\rva)}{d^{\star }_{\rvs}} \right)  \label{eq:pf_of_redcution_convex_TV}\\
        & =\sum_{\rvs \in \mathsf{S}} d^{\star}_{\rvs}\overline{\TV}_{\rvs}(\rva_{\rvs}^{\star}) = \sum_{\rvs \in \mathsf{S}} \int_{\mathsf{A}(\rvs)} \overline{\TV}_{\rvs}(\rva)d^{\star}_{\rvs} \delta_{\rva_{\rvs}^{\star}}(\mathrm{d}\rva)   \\
        & = \sum_{\rvs \in \mathsf{S}} \int_{\mathsf{A}(\rvs)} \overline{\TV}_{\rvs}(\rva) d^\star(\rvs, \mathrm{d}\rva) 
    \end{align}
    Since unnormalized total variation divergence is convex, inequality (\ref{eq:pf_of_redcution_convex_TV}) follows from the Jensen inequality of convex function in integral form.
    \item Objective function: 
    We need to show that $d^{\star}(\rvs,\cdot)$ achieves a value equal to or exceeding that of $\tilde{d}(\rvs,\cdot)$ in the context of the objective function. We adopt a similar approach as before: use Jensen’s inequality for concave functions and apply it to (\ref{eq:pf_of_reduction_concave_TV}). The case $d_\rvs^{\star} = 0$ is excluded, since it makes no contribution to the objective.
    \begin{align}
        & \sum_{\rvs \in \mathsf{S}} \int_{\mathsf{A}(\rvs)} H(\rva) \tilde{d}(\rvs,\mathrm{d}\rva) =  \sum_{\rvs \in \mathsf{S}} d_\rvs^{\star} \int_{\mathsf{A}(\rvs)} H(\rva) \frac{\tilde{d} (\rvs,\mathrm{d}\rva)}{d^{\star }_{\rvs}} \\
        \leq & \sum_{\rvs \in \mathsf{S}} d_\rvs^{\star} H({\int_{\mathsf{A}(\rvs)} \rva \frac{\tilde{d} (\rvs,\mathrm{d}\rva)}{d^{\star }_{\rvs}}}) \label{eq:pf_of_reduction_concave_TV} \\
        = & \sum_{\rvs \in \mathsf{S}} d_\rvs^{\star} H({\rva^{\star}_\rvs})
        = \sum_{\rvs \in \mathsf{S}} \int_{\mathsf{A}(\rvs)} d_\rvs^{\star} \delta_{\rva^\star_\rvs}(\mathrm{d} \rva)  H(\rva) = \sum_{\rvs \in \mathsf{S}} \int_{\mathsf{A}(\rvs)} d^{\star} (\rvs,\mathrm{d} \rva) H(\rva)
    \end{align}
\end{enumerate}
For $d_\rvs^\star>0$, the policy obtained from the occupancy measure is
\begin{align}
    \pi(\mathrm{d} \rva \mid \rvs)
    =
    \delta_{\rva_\rvs^\star}(\mathrm{d} \rva).
\end{align}
If $d_\rvs^\star=0$, the policy at state $\rvs$ can be chosen arbitrarily, and
we choose the same deterministic policy
$\delta_{\rva_\rvs^\star}$. 

This means the policy chooses the unique action $\rva_s^{\star}\in \mathsf{A}(\rvs)$ with probability one. Replacing $d(\rvs,\cdot)$ in $\mathbf{P1}$ by $d^{\star}(\rvs,\cdot)$ gives $\mathbf{P2}$ with an objective function value equal or higher. Conversely, any feasible $(d_\rvs,\rva_\rvs)$ for $\mathbf{P2}$ induces a feasible $d(\rvs,\mathrm{d}\rva)=d_\rvs\delta_{\rva_\rvs}(\mathrm{d}\rva)$ for $\mathbf{P1}$ with the same objective. Hence, $\mathbf{P1}$ and $\mathbf{P2}$ are equivalent, and $\mathbf{P1}$ reduces the infinite-dimensional decision variable to the finite set $\{\rva_\rvs^{\star},\, d_\rvs^{\star}\}$.

\section{Proof of Theorem 2 and Theorem 3} \label{pf:thm2}

To solve the convex optimization problem $\mathbf{P3\text{-}(2,1)}$, we consider the KKT conditions. However, the unnormalized total variation constraint poses a challenge as it is not differentiable when $ \frac{x_{0,0}}{d_0} = p_{0,0}$ and $ \frac{x_{1,0}}{d_1} = p_{1,0}$, preventing us from performing a direct partial differentiation on this inequality constraint. Therefore, we introduce new variables $c_{0,0}, c_{1,0}$ such that $d_0 |\frac{x_{0,0}}{d_0} - p_{0,0}| = |x_{0,0} - d_0 p_{0,0}| \leq c_{0,0}$ and $d_1|\frac{x_{1,0}}{d_1} - p_{1,0}| = |x_{1,0} - d_1 p_{1,0}|\leq c_{1,0}$ to replace the absolute function. $c_{0,0}, c_{1,0}$ serve as the parameters defining the upper bound of unnormalized total variation between two states. We also note that, in the subsequent analysis, we assume $0<p_{s,0}<1$ for every $s\in\{0,1\}$, so that the transition probabilities are not degenerate. Moreover, throughout this proof we focus on the case where $d_0, d_1$ are nonzero, because if they are zero, the optimal actions $a_{0,0}, a_{0,1}$ may be chosen arbitrarily, as mentioned in Section \ref{sec:formulation}.  

The optimization problem $\mathbf{P3\text{-}(2,1)}$ can be reformulated as follows. 
\begin{align}
    \max_{x_{0,0},x_{1,0}, d_0, d_1, c_{0,0}, c_{1,0}} & ~ d_0 H(\frac{x_{0,0}}{d_0}) + d_1 H(\frac{x_{1,0}}{d_1})  \label{eq:nx0_obj_func_TV}\\
    \textnormal{subject to } 
    & ~ c_{0,0} + c_{1,0}\leq \frac{b}{2} \label{eq:nx0_constraint_1_TV}\\ 
    & -c_{0,0} \leq x_{0,0} - d_0 p_{0,0}\leq c_{0,0} \label{eq:nx0_constraint_2_TV}\\
    & -c_{1,0} \leq x_{1,0} - d_1 p_{1,0} \leq c_{1,0} \label{eq:nx0_constraint_3_TV}\\
    & ~  d_0 -\gamma x_{0,0}  - \gamma x_{1,0} = \mathbf{d}^{\gamma}_0 \label{eq:nx0_constraint_4_TV}\\
    & ~ d_0 + d_1 = 1 \label{eq:nx0_constraint_5_TV}\\
    & ~ 0 \leq x_{0,0}  \leq d_0 ,  ~ 0 \leq x_{1,0}  \leq d_1 \label{eq:nx0_constraint_6_TV}
\end{align}
The Lagrangian function of this problem is: 
\begin{align}
    \mathcal{L} = & - \left( d_0 H(\frac{x_{0,0}}{d_0}) + d_1 H(\frac{x_{1,0}}{d_1}) \right) + \lambda \left( c_{0,0} + c_{1,0} - \frac{b}{2} \right) \nonumber \\
    & \alpha_0 (-c_{0,0} -x_{0,0} +d_0 p_{0,0}) + \beta_0 (x_{0,0} - d_0 p_{0,0} -c_{0,0} ) \nonumber \\
    & + \alpha_1 (-c_{1,0} -x_{1,0} +d_1 p_{1,0}) + \beta_1 (x_{1,0} - d_1 p_{1,0} -c_{1,0} ) \nonumber \\
    & + \omega_0 \left( d_0 - \mathbf{d}^{\gamma}_0 -\gamma (x_{0,0} +x_{1,0})\right) + \omega_1 \left( d_0 +d_1 -1 \right) \nonumber \\
    & - \nu_0 x_{0,0} - \nu_1 x_{1,0} + \mu_0(x_{0,0} -d_0) + \mu_1 (x_{1,0} - d_1)
\end{align}
where $\lambda, \alpha_s, \beta_s ,\omega_s,  \nu_s, \mu_s$ are the Lagrange multipliers of constraints (\ref{eq:nx0_constraint_1_TV}) to (\ref{eq:nx0_constraint_6_TV}).
The KKT conditions are written as follows: 
\begin{enumerate}
    \item Stationarity: 
    \begin{align}
            \frac{\partial  \mathcal{L}}{\partial x_{0,0}} & = - \log\left( \frac{d_0 - x_{0,0}}{x_{0,0}} \right) - \alpha_0 + \beta_0 - \gamma \omega_0 - \nu_0 + \mu_0 = 0 \label{eq:stationary1} \\
            \frac{\partial  \mathcal{L}}{\partial x_{1,0}} & = - \log\left( \frac{d_1 - x_{1,0}}{x_{1,0}} \right) - \alpha_1 + \beta_1 - \gamma \omega_0 - \nu_1 + \mu_1 = 0 \label{eq:stationary2} \\
            \frac{\partial  \mathcal{L}}{\partial d_{0}} & = - H(\frac{x_{0,0}}{d_0}) + \frac{x_{0,0}}{d_0} \log \left( \frac{d_0-x_{0,0}}{x_{0,0}}\right) + \alpha_0 p_{0,0} - \beta_0 p_{0,0} + \omega_0 + \omega_1 - \mu_0 = 0\\
            \frac{\partial  \mathcal{L}}{\partial d_{1}} & = - H(\frac{x_{1,0}}{d_1}) + \frac{x_{1,0}}{d_1} \log \left( \frac{d_1-x_{1,0}}{x_{1,0}}\right) +\alpha_1 p_{1,0} - \beta_1 p_{1,0} + \omega_1 - \mu_1= 0 \\
            \frac{\partial  \mathcal{L}}{\partial c_{0,0}} & = \lambda - \alpha_0 - \beta_0
            = 0 \\
            \frac{\partial  \mathcal{L}}{\partial c_{1,0}} & = \lambda - \alpha_1 - \beta_1 = 0 \label{eq:stationary6}
    \end{align}
    \item Primal feasibility: 
        \begin{align}
            & c_{0,0} + c_{1,0} \leq \frac{b}{2} \label{eq:primal_feas1}\\
            & -c_{0,0} -x_{0,0} + d_0 p_{0,0} \leq 0, ~ -c_{1,0} -x_{1,0} + d_1 p_{1,0} \leq 0 \\
            & x_{0,0} - d_0 p_{0,0}  - c_{0,0} \leq 0, ~ x_{1,0} - d_1 p_{1,0} - c_{1,0} \leq 0 \\
            & d_0 = \mathbf{d}^{\gamma}_0 + \gamma (x_{0,0} +x_{1,0}) \\
            & d_0 + d_1 = 1 \\
            & x_{0,0} \geq 0, ~ x_{1,0} \geq 0 \\
            & x_{0,0} - d_0 \leq 0, ~ x_{1,0} - d_1 \leq 0  \label{eq:primal_feas7}
        \end{align}
    \item Dual feasibility:
        \begin{align}
                & \lambda \geq 0 \label{eq:dual_feas1}\\
                & \alpha_0 \geq 0, ~ \alpha_1 \geq 0 \label{eq:dual_feas2}\\
                & \beta_0 \geq 0, ~ \beta_1 \geq 0 \label{eq:dual_feas3}\\
                & \nu_0 \geq 0, ~ \nu_1 \geq 0 \label{eq:dual_feas4}\\
                & \mu_0 \geq 0, ~ \mu_1 \geq 0 \label{eq:dual_feas5}
        \end{align}
    \item Complementary slackness:
    \begin{align}
                & \lambda \left( c_{0,0} + c_{1,0} - \frac{b}{2}\right) = 0 \label{eq:complementary1}\\
                & \alpha_0 (-c_{0,0} -x_{0,0} + d_0 p_{0,0}) = \alpha_1 (-c_{1,0} -x_{1,0} + d_1 p_{1,0}) =  0 \\
                & \beta_0 (x_{0,0} - d_0 p_{0,0} -c_{0,0}) = \beta_1 (x_{1,0} - d_1 p_{1,0} -c_{1,0}) = 0 \\
                & \nu_0 x_{0,0} = \nu_1 x_{1,0} = 0 \\
                & \mu_0(x_{0,0} - d_0)= \mu_1( x_{1,0} - d_1) = 0 \label{eq:complementary7}
    \end{align}
\end{enumerate}
Solutions to the KKT conditions will be presented in several cases, depending on the values of $p_{0,0}$ and $p_{1,0}$, separated into different regimes by the cost constraint $b$. The expressions for the primal variables $x_{s,0}, d_s$ and the dual variables $\omega_s, \nu_s, \mu_s$ where $s \in \{0,1\}$ are identical across all scenarios, as presented below.
\begin{align}
        & x_{0,0} = a_{0,0} d_0 \label{eqn:x0}\\
        & x_{1,0} = a_{1,0} d_1 \\
        & d_0 = \frac{\mathbf{d}^{\gamma}_0 + \gamma a_{1,0}}{1 - \gamma a_{0,0} + \gamma a_{1,0}} \\
        & d_1 = 1 - d_0 \label{eqn:d1}\\
        & \omega_0 = \frac{-1}{\gamma} \log(\frac{1-a_{0,0}}{a_{0,0}}) - \frac{1}{\gamma} \alpha_0 + \frac{1}{\gamma} \beta_0 = \frac{-1}{\gamma} \log(\frac{1-a_{1,0}}{a_{1,0}}) - \frac{1}{\gamma} \alpha_1 + \frac{1}{\gamma} \beta_1 \label{eqn:w0}\\
        & \omega_1 = -\log(1-a_{1,0}) - \alpha_1 p_{1,0} + \beta_1 p_{1,0} \\
        & \nu_s = \mu_s = 0, ~ s \in \{ 0, 1\}\label{eqn:nu}
\end{align}
Therefore, we will not write them out particularly in each case, since they can derived from $a_{s,0}, \alpha_s, \beta_s, ~ s \in \{0,1\} $. 

We define the following functions for simplicity:
\begin{align} 
    & L_s = \log \left( \frac{1-a_{s,0}}{a_{s,0}} \right), \quad \forall s \in \{0,1 \} \\
    & z_0(a_{0,0}, a_{1,0})  =  \left( -1-\gamma p_{1,0}\right)  L_0  +\gamma p_{1,0}L_1+\gamma \log \left( \frac{1-a_{0,0}}{1-a_{1,0}} \right)\\
    & z_1(a_{0,0}, a_{1,0})  =  \left( -\gamma p_{0,0}\right)  L_0 +(-1+ \gamma p_{0,0} ) L_1 +\gamma \log \left( \frac{1-a_{0,0}}{1-a_{1,0}} \right).
\end{align}
Thus, the value of $z_0+z_1$ and $z_1 - z_0$ become
\begin{align}
    & (z_0 +z_1)(a_{0,0}, a_{1,0}) = (-1-\gamma p_{1,0} - \gamma p_{0,0})L_0 + (-1 + \gamma p_{1,0} + \gamma p_{0,0}) L_1 +2\gamma \log \left( \frac{1-a_{0,0}}{1-a_{1,0}} \right) \\
    & (z_1-z_0)(a_{0,0}, a_{1,0}) = \frac{1}{M}(L_0 - L_1).
\end{align}

\subsection{Proof of Theorem 2}
In Theorem \ref{thm:p0p1sameside}, the assumption is that $p_{0,0}, p_{1,0} \geq \frac{1}{2}$. There are three regimes:
    \begin{enumerate}
        \item The regime $b \in [0, b_1) $:       
        In this regime, the two cases  $p_{1,0}\geq p_{0,0}\geq \frac{1}{2}$ and $p_{0,0}> p_{1,0}\geq \frac{1}{2}$ lead to different solution expressions, and let us consider the first case for now. With the solution of $a_{0,0},a_{1,0}$ given in the theorem, and in the given range of $b$, the range of $a_{0,0}$ and $a_{1,0}$ are 
        \begin{align}
                a_{0,0} = p_{0,0}, ~ a_{1,0} \in [p_{0,0}, p_{1,0}],
        \end{align}
        The variable assignments are as follows. 
        \begin{enumerate}
        \item Primal variable assignments: 
        \begin{align}
            & a_{0,0} = p_{0,0},\quad a_{1,0} = \eta_1^{-}(b)\\            &    c_{0,0} = 0, \quad c_{1,0} = \frac{b}{2},      
        \end{align}
        Using (\ref{eqn:x0})-(\ref{eqn:d1}), all the primal variables are assigned. Note that $x_{0,0}$ and $x_{1,0}$ are non-zero in this setting because $a_{0,0}$ and $a_{1,0}$ are not zero. As a result, we obtain the stationary conditions (\ref{eq:stationary1}) and (\ref{eq:stationary2}). 
        \item Dual variable assignments:         
        \begin{align}
                 & \alpha_0 = \tfrac{\lambda + Mz_0}{2} \\
                 & \beta_0 = \tfrac{\lambda - Mz_0}{2} \\
                 & \alpha_1 = Mz_1 = \tfrac{\lambda + Mz_1}{2} \\
                 & \beta_1 = \tfrac{\lambda - Mz_1}{2} = 0 \\
                 & \lambda = \alpha_1 
        \end{align}        
        With (\ref{eqn:w0})-(\ref{eqn:nu}), all the dual variables are assigned. 
        \end{enumerate}
       The assignments satisfy the KKT conditions. With the primal and dual variables displayed above, it is straightforward to show that (\ref{eq:stationary1})-(\ref{eq:stationary6}), (\ref{eq:primal_feas1})-(\ref{eq:primal_feas7}), (\ref{eq:dual_feas4})-(\ref{eq:dual_feas5}), and (\ref{eq:complementary1})-(\ref{eq:complementary7}) are satisfied.         
        It remains to verify (\ref{eq:dual_feas1})-(\ref{eq:dual_feas3}). It suffices to prove that $\lambda \geq |Mz_0|$; this will then imply that $\alpha_s, \beta_s,$ and $\lambda$ are all nonnegative.

        Here, since $\lambda=\alpha_1=Mz_1$, proving that $\lambda \ge |Mz_0|$ reduces to proving that $z_1 \ge |z_0|$. Equivalently, it suffices to prove the two inequalities $z_1 - z_0 \ge 0$ and $z_1 + z_0 \ge 0$. The inequality $z_1 - z_0 \ge 0$ follows readily, since in the case $a_{1,0} \ge a_{0,0} = p_{0,0} \ge \frac{1}{2}$ we have $L_0 > L_1$. Therefore, it remains only to establish that $z_1 + z_0 \ge 0$ under $a_{1,0} \ge a_{0,0} = p_{0,0} \ge \frac{1}{2}$.

        The argument below establishes that $z_1+z_0$ is increasing in $a_{1,0}$ by computing $\frac{\partial (z_1+z_0)}{\partial a_{1,0}}$:
    \begin{align}
    \frac{\partial (z_1+z_0)}{\partial a_{1,0}} & = (-1+\gamma p_{1,0} + \gamma p_{0,0}) \frac{-1}{a_{1,0}(1-a_{1,0})} + 2 \gamma \frac{1}{(1-a_{1,0})} \notag \\
    & = \frac{1}{1-a_{1,0}}\left(1 + \gamma (2a_{1,0} - p_{1,0}-p_{0,0}) \right).
    \end{align}
    This derivative is  nonnegative, since $a_{1,0} \ge \frac{1}{2}$ and $p_{0,0}+p_{1,0} \le 2$, we obtain
    $$1 + \gamma (2a_{1,0} - p_{1,0}-p_{0,0}) \ge 1-\gamma \ge 0.$$

    Furthermore, the minimum of $z_1+z_0$ is attained at $a_{1,0}=a_{0,0}=p_{0,0}$. We verify below that $(z_1+z_0)(p_{0,0},p_{0,0}) \ge 0$:
    \begin{align}
    z_1+z_0(p_{0,0},p_{0,0})
    &= (-1-\gamma p_{1,0}-\gamma p_{0,0}-1+\gamma p_{1,0}+\gamma p_{0,0})
    \log\left(\frac{1-p_{0,0}}{p_{0,0}}\right) \\
    &= -2\log\left(\frac{1-p_{0,0}}{p_{0,0}}\right) \ge 0,\qquad \text{when } p_{0,0}\ge \frac12.
    \end{align}
    Therefore, under the conditions $a_{1,0}\ge a_{0,0}=p_{0,0}\ge \frac12$, we conclude that $z_1+z_0\ge 0$, which establishes that all $\alpha_s$, $\beta_s$, and $\lambda$ are nonnegative.

        Now let us turn our attention to the case when $p_{0,0} \geq p_{1,0} \geq \frac{1}{2}$. We can determine the range of $a_{0,0}$ and $a_{1,0}$ as 
        \begin{align}
                a_{0,0} \in [p_{1,0},p_{0,0}], ~ a_{1,0} = p_{1,0} .
        \end{align}
        The variable assignments are as follows. 
        \begin{enumerate}
        \item Primal variable assignments: 
        \begin{align}
                & a_{0,0} = \eta_0^{+}(b),\quad a_{1,0} = p_{1,0} \\
                & c_{0,0} = \frac{b}{2},~  \quad c_{1,0} = 0 
        \end{align}
        Using (\ref{eqn:x0})-(\ref{eqn:d1}), all the primal variables are assigned. Observe that $x_{0,0}$ and $x_{1,0}$ will not be zero here as well. 
        \item Dual variable assignments:
        \begin{align}
                & \alpha_0 = Mz_0 = \tfrac{\lambda + Mz_0}{2} \\
                & \beta_0 = \tfrac{\lambda - Mz_0}{2} = 0\\
                & \alpha_1 = \tfrac{\lambda + Mz_1}{2} \\
                & \beta_1 = \tfrac{\lambda - Mz_1}{2} \\
                & \lambda = \alpha_0         
        \end{align}
        With (\ref{eqn:w0})-(\ref{eqn:nu}), all the dual variables are assigned.
        \end{enumerate}
        The rest of the proof follows the same argument as in the previous case to verify that these assignments satisfy the KKT conditions; therefore, we omit the details to avoid repetition.

    \item $b \in [b_1, b_2)$: In this regime, we have the following assignments.
     \begin{enumerate}    
    \item Primal variable assignments: 
    \begin{align}
            & a_{0,0} = a_{1,0} = -M(\frac{b}{2} - \mathbf{d}_0^{\gamma}p_{0,0} - \mathbf{d}_1^{\gamma}p_{1,0} - \gamma p_{1,0})\\
            & c_{0,0} = d_0(p_{0,0}-a_{0,0}),\quad c_{1,0} = d_1(p_{1,0}-a_{1,0}).
    \end{align}
    Using (\ref{eqn:x0})-(\ref{eqn:d1}), all the primal variables are assigned. Observe that $x_{0,0}$ and $x_{1,0}$ will not be zero here, indicating that stationary conditions can be derived.  
    \item Dual variable assignments: 
    \begin{align}
        \begin{cases}
        & \alpha_0 = \alpha_1 = M \log \left( \frac{a_{0,0}}{1-a_{0,0}} \right)  \\
        & \beta_0 = \beta_1 = 0\\
        & \lambda = \alpha_1 \\
        \end{cases}
    \end{align}
    \end{enumerate}
    With (\ref{eqn:w0})-(\ref{eqn:nu}), all the dual variables are assigned. 
    
    Similar to the previous discussion, we wish to show that the assignments satisfy all the KKT conditions. With the primal and dual variables displayed above, one can conveniently confirm that (\ref{eq:stationary1})-(\ref{eq:stationary6}), (\ref{eq:primal_feas1})-(\ref{eq:primal_feas7}), (\ref{eq:dual_feas3})-(\ref{eq:dual_feas5}), and (\ref{eq:complementary1})-(\ref{eq:complementary7}) are satisfied by substituting these variables. Since $a_{0,0} \geq \frac{1}{2}$, we have that $\alpha_s, \lambda$ are non-negative, leading to   (\ref{eq:dual_feas1}) and (\ref{eq:dual_feas2}) in this case.
    \item $b \in [b_2, \infty)$: In this regime, we have  
    \begin{enumerate}
    \item Primal variables: 
    \begin{align}
        & a_{0,0} = a_{1,0} = \frac{1}{2}\\
        & c_{0,0} = d_0(p_{0,0}-\frac{1}{2}), \quad c_{1,0} = d_1(p_{1,0}-\frac{1}{2}).
    \end{align}
    Using (\ref{eqn:x0})-(\ref{eqn:d1}), all the primal variables are assigned. Observe that $x_{0,0}$ and $x_{1,0}$ will not be zero here as well. 
    \item Dual variables: 
    \begin{align}
        & \alpha_0 = \alpha_1 = \beta_0 = \beta_1  = \lambda = 0.
    \end{align}
    All the dual variables are assigned using (\ref{eqn:w0})-(\ref{eqn:nu}). 
    \end{enumerate}
    With the primal and dual variables displayed above, one can conveniently confirm that (\ref{eq:stationary1})-(\ref{eq:stationary6}), (\ref{eq:primal_feas1})-(\ref{eq:primal_feas7}), (\ref{eq:dual_feas1})-(\ref{eq:dual_feas5}), and (\ref{eq:complementary1})-(\ref{eq:complementary7}) are satisfied by substituting these variables. 
\end{enumerate}

\subsection{Proof of Theorem 3} \label{pf:thm3}
Theorem \ref{thm:p0p1diffside} applies in the two settings $p_{0,0} \leq \frac{1}{2} \leq p_{1,0}$ and $p_{0,0} \geq \frac{1}{2} \geq p_{1,0}$.We establish the first case by applying the KKT conditions given below. In the second case, the solution is derived by adjusting in the opposite direction, and the KKT verification proceeds in the same way. We thus omit the details.

Under $p_{0,0} \leq \frac{1}{2} \leq p_{1,0}$ case, there are again three regimes:
\begin{enumerate}
    \item The regime $b \in [0, b_1^{-}) $: In this regime, the two cases $|\frac{1}{2}-p_{1,0}| \geq |\frac{1}{2}-p_{0,0}|$ and $|\frac{1}{2}-p_{1,0}| < |\frac{1}{2}-p_{0,0}|$ lead to different solution expressions, and let us consider the first case for now. Observe that the first case has $p_{1,0} + p_{0,0} -1 \geq 0$, then $b_1^{-}$ in this case can be written as: 
    \begin{align}
        &b_1^{-}  = 2\left( \frac{1-\gamma p_{0,0}-\mathbf{d}^{\gamma}_0 }{1-\gamma p_{0,0}+\gamma \psi_{1} } \right)  (p_{1,0}-\psi_{1} )
    \end{align}
    by (\ref{eq:bl'_def}). 

    With the solution of $a_{0,0},a_{1,0}$ given in the theorem, and in the given range of $b$, the range of $a_{0,0}$ and $a_{1,0}$ are
    \begin{align}
        a_{0,0} = p_{0,0}, ~ a_{1,0} \in [\psi_1, p_{1,0}]\subset[\frac{1}{2}, p_{1,0}]
    \end{align}
    The variable assignments are as follows. 
    \begin{enumerate}
        \item Primal variable assignments:
        \begin{align}
            & a_{0,0} = p_{0,0},\quad a_{1,0} = \eta_1^{-}(b) \\
            & c_{0,0} = 0, \quad c_{1,0} = \frac{b}{2}
        \end{align}
         Using (\ref{eqn:x0})-(\ref{eqn:d1}), all the primal variables are assigned. Note that in this scenario, $x_{0,0}$ and $x_{1,0}$ are non-zero, enabling us to address the stationarity conditions in (\ref{eq:stationary1}) and (\ref{eq:stationary2}).
        \item Dual variable assignments: 
        \begin{align}
            & \alpha_0 = \tfrac{\lambda+Mz_0}{2}= \tfrac{M}{2}(z_1+z_0) = \tfrac{M}{2} (\Phi_{-}(a_{1,0}) - \Phi_{+}(p_{0,0})) \\
            & \alpha_1 = \tfrac{\lambda+Mz_1}{2} = Mz_1\\
            & \beta_0 = \tfrac{\lambda-Mz_0}{2} = \tfrac{M}{2}(z_1- z_0) \\
            & \beta_1 = \tfrac{\lambda-Mz_1}{2} = 0\\
            & \lambda =\alpha_1 
        \end{align}
        With (\ref{eqn:w0})-(\ref{eqn:nu}), all the dual variables are assigned.
    \end{enumerate}
    We show that the assignments satisfy the KKT conditions. With the primal and dual variables displayed above, one can conveniently confirm that (\ref{eq:stationary1})-(\ref{eq:stationary6}), (\ref{eq:primal_feas1})-(\ref{eq:primal_feas7}), (\ref{eq:dual_feas4})-(\ref{eq:dual_feas5}), and (\ref{eq:complementary1})-(\ref{eq:complementary7}) are satisfied by substituting these variables. It remains to verify  (\ref{eq:dual_feas1})-(\ref{eq:dual_feas3}). As we argued earlier in the proof of Theorem \ref{thm:p0p1sameside}, it suffices to verify that $z_1 - z_0 \ge 0$ and $z_1 + z_0 \ge 0$ in order to conclude that $\alpha_s$, $\beta_s$, and $\lambda$ are nonnegative.

    One can readily verify that \(z_1 - z_0 = \frac{1}{M}(L_0 - L_1) \ge 0\), because in the situation \(a_{1,0} \ge \frac{1}{2} \ge a_{0,0} = p_{0,0}\) we have \(L_0 > 0\) and \(L_1 < 0\).
    
        To demonstrate $z_0 +z_1 \geq 0$, it is necessary to establish that $ (\Phi_{-}(a_{1,0}) - \Phi_{+}(p_{0,0}))$ is non-negative. Employing a similar approach as earlier, we first show that $(\Phi_{-}(a_{1,0}) - \Phi_{+}(p_{0,0}))$ increases when $a_{1,0} \in [\psi_1, p_{1,0}]$ by way of (\ref{eq:psia1_da1_geq0}). Then (\ref{eq:case3a_rho0_psi1_geq0}) confirms that its minimum value is zero, occurring at $a_{1,0} = \psi_1$.
    \begin{align}
        \frac{\partial}{\partial a_{1,0}}  (\Phi_{-}(a_{1,0}) - \Phi_{+}(p_{0,0}))& = \Phi_-'(a_{1,0}) =\frac{2\gamma a_{1,0}+1-\gamma p_{0,0} -\gamma p_{1,0}}{(1-a_{1,0})a_{1,0}} \geq 0 \label{eq:psia1_da1_geq0}\\
         (z_0+z_1)(p_{0,0}, a_{1,0} = \psi_1)& = (\Phi_{-}(\psi_1) - \Phi_{+}(p_{0,0}))  = 0 \label{eq:case3a_rho0_psi1_geq0} 
    \end{align}
    The inequality in (\ref{eq:psia1_da1_geq0}) follows from the fact that $a_{1,0} \geq \frac{1}{2}$. 
    The above establishes that $\alpha_s, \beta_a, \lambda$ are all non-negative.
    
     Now let us turn our attention to the case when $|\frac{1}{2}-p_{1,0}| <|\frac{1}{2}-p_{0,0}| $, implying $1-p_{1,0} - p_{0,0} \geq 0$. $b_1^{-}$ in this case can be written as: 
    \begin{align}
        &b_1^{-}  = 2\left( \frac{\mathbf{d}^{\gamma}_0 +\gamma p_{1,0}}{1-\gamma \psi_{0} +\gamma p_{1,0}} \right)  (\psi_{0} -p_{0,0})
    \end{align}
    again by (\ref{eq:bl'_def}).  
    
     We can determine the range of $a_{0,0}$ and $a_{1,0}$ as 
     \begin{align}
        a_{0,0} \in [p_{0,0},\psi_0], ~ a_{1,0} = p_{1,0}.
    \end{align}
    The variable assignments are as follows. 
    \begin{enumerate}
        \item Primal variable assignments: 
        \begin{align}
            & a_{0,0} = \eta_0^{-}(b), ~ a_{1,0} = p_{1,0} \\
            & c_{0,0} = \frac{b}{2}, ~ c_{1,0} = 0 
        \end{align}
        Using (\ref{eqn:x0})-(\ref{eqn:d1}), all the primal variables are assigned. Note that $x_{0,0}$ and $x_{1,0}$ are also non-zero in this context, allowing us to derive stationary conditions (\ref{eq:stationary1}) and (\ref{eq:stationary2}).
        \item Dual variable assignments:
        \begin{align}
            & \alpha_0 = \tfrac{\lambda +Mz_0}{2} = 0 \\
            & \alpha_1 = \tfrac{\lambda +Mz_1}{2} = \tfrac{M}{2}(z_1-z_0) \\
            & \beta_0 =\tfrac{\lambda-Mz_0}{2}=  -Mz_0  \\
            & \beta_1 = \tfrac{\lambda -Mz_1}{2} = \tfrac{M}{2} (-z_0 - z_1) = \tfrac{M}{2}(\Phi_{+}(a_{0,0})- \Phi_{-}(p_{1,0}))  \\
            & \lambda = \beta_0
        \end{align}
        With (\ref{eqn:w0})-(\ref{eqn:nu}), all the dual variables are assigned.
    \end{enumerate}
    We wish to show that the assignments satisfy all the KKT conditions. With the primal and dual variables displayed above, one can conveniently confirm that (\ref{eq:stationary1})-(\ref{eq:stationary6}), (\ref{eq:primal_feas1})-(\ref{eq:primal_feas7}), (\ref{eq:dual_feas4})-(\ref{eq:dual_feas5}), and (\ref{eq:complementary1})-(\ref{eq:complementary7}) are satisfied by substituting these variables. It remains to verify (\ref{eq:dual_feas1})-(\ref{eq:dual_feas3}). As in the previous case where we proved nonnegativity, it suffices to verify that $z_1 - z_0 \ge 0$ and $-z_1 - z_0 \ge 0$. 

    One can readily verify that \(z_1 - z_0 = \frac{1}{M}(L_0 - L_1) \ge 0\), because in the situation \(a_{1,0} = p_{1,0} \ge \frac{1}{2} \ge a_{0,0} \) we have \(L_0 > 0\) and \(L_1 < 0\).

    For $-z_0-z_1 \ge 0$, a similar procedure can be used to show its non-negativity. (\ref{eq:case3a_rho1decreasing}) implies that $M(\Phi_{+}(a_{0,0})- \Phi_{-}(p_{1,0}))$ is monotonically non-increasing in $a_{0,0} \in [p_{0,0}, \psi_0]$, given that $a_{0,0} \leq \frac{1}{2}$. Furthermore, as shown in (\ref{eq:case3a_rho1_psi0_geq0}), the minimum value of $M(\Phi_{+}(a_{0,0})- \Phi_{-}(p_{1,0}))$ over the interval $a_{0,0} \in [p_{0,0}, \psi_0]$ is zero, which is derived from the definition of $\psi_0$ in Theorem \ref{thm:p0p1diffside}.  
    \begin{align}
        & \frac{\partial }{\partial a_{0,0}} (\Phi_{+}(a_{0,0})- \Phi_{-}(p_{1,0})) = \Phi^{\prime}_+ (a_{0,0}) \notag \\
        & = \frac{2\gamma a_{0,0} - 1 - \gamma p_{0,0} - \gamma p_{1,0}}{(1-a_{0,0})a_{0,0}} \leq 0, \forall a_{0,0} \in [p_{0,0},  \psi_0] \subset[p_{0,0}, \frac{1}{2}] \label{eq:case3a_rho1decreasing} 
    \end{align} 
    \begin{align}
          (-z_0-z_1)(a_{0,0}=\psi_0, p_{1,0}) = (\Phi_{+}(\psi_0)- \Phi_{-}(p_{1,0})) =0  \label{eq:case3a_rho1_psi0_geq0}
    \end{align}
    Therefore (\ref{eq:dual_feas1})-(\ref{eq:dual_feas3}) hold in this case. 

    \item $b \in [b_1^{-}, b_2^{-}]$: Let us first define the following set
    \begin{align}
    &S_b^{-} \triangleq  \bigg\{ (a_{0,0},a_{1,0}) \in [p_{0,0},\frac{1}{2}]\times[\frac{1}{2},p_{1,0}] ~ |\nonumber \\
            &\qquad\quad \Phi_-(a_{1,0}) = \Phi_+(a_{0,0})  \mbox{~and~} m^{-}(a_{0,0},a_{1,0}) = \frac{b}{2} \bigg\}\label{eqn:Sb}
    \end{align}
    It will be shown shortly that the set $S_b^-$ is a singleton set in this regime. We have the following assignments. 
    \begin{enumerate}
        \item Primal variable assignments: 
        \begin{align}
            &(a_{0,0}, a_{1,0}) \in S_b^{-} \\
            &c_{0,0} = d_0(a_{0,0}-p_{0,0}), \quad c_{1,0} = d_1(p_{1,0}-a_{1,0}) 
        \end{align} 
         Using (\ref{eqn:x0})-(\ref{eqn:d1}), all the primal variables are assigned. Note that $x_{0,0}$ and $x_{1,0}$ will not be zero here as well, thus allowing the derivation of stationary conditions. 
        \item Dual variable assignments: 
        \begin{align}
            & \alpha_0 = 0\\
            & \alpha_1 = \frac{1}{2}\log \left( \frac{1-a_{0,0}}{a_{0,0}} \right) -\frac{1}{2}\log \left( \frac{1-a_{1,0}}{a_{1,0}} \right) \\
            & \beta_0 = \alpha_1 \\
            & \beta_1 = 0 \\
            & \lambda = \alpha_1
        \end{align}
        With (\ref{eqn:w0})-(\ref{eqn:nu}), all the dual variables are assigned. We can confirm that (\ref{eq:stationary1})-(\ref{eq:stationary6}), (\ref{eq:primal_feas1})-(\ref{eq:primal_feas7}), (\ref{eq:dual_feas4})-(\ref{eq:dual_feas5}), and (\ref{eq:complementary1})-(\ref{eq:complementary7}) are satisfied by substituting these variables. According to the definition of $S_b^{-}$, if $(a_{0,0}, a_{1,0})$ is an element in $S_b^{-}$, then it follows that $a_{0,0} \leq \frac{1}{2} \leq a_{1,0}$. Clearly, the conditions (\ref{eq:dual_feas1})-(\ref{eq:dual_feas3}) remain valid since $a_{0,0} \leq \frac{1}{2} \leq a_{1,0}$.
    \end{enumerate}
    
    \item $b \in (b_2^{-}, \infty)$: In this regime, we have 
    \begin{enumerate}
    \item Primal variables: 
    \begin{align}
        & a_{0,0} = a_{1,0} = \frac{1}{2}\\
        & c_{0,0} = d_0(\frac{1}{2}-p_{0,0}), \quad c_{1,0} = d_1(p_{1,0}-\frac{1}{2}).
    \end{align}
    Using (\ref{eqn:x0})-(\ref{eqn:d1}), all the primal variables are assigned. Observe that $x_{0,0}$ and $x_{1,0}$ will not be zero here as well. 
    \item Dual variables: 
    \begin{align}
        & \alpha_0 = \alpha_1 = \beta_0 = \beta_1  = \lambda = 0.
    \end{align}
    All the dual variables are assigned using (\ref{eqn:w0})-(\ref{eqn:nu}). 
    \end{enumerate}
    With the primal and dual variables displayed above, we confirm that (\ref{eq:stationary1})-(\ref{eq:stationary6}), (\ref{eq:primal_feas1})-(\ref{eq:primal_feas7}), (\ref{eq:dual_feas1})-(\ref{eq:dual_feas5}), and (\ref{eq:complementary1})-(\ref{eq:complementary7}) are satisfied by substituting these variables. 
\end{enumerate}

\subsection{Proof That $S_b^-$ Is a Singleton Set} 
\label{sec:appendix_sbnonempty}

\begin{figure}[t!]
    \centering
    \includegraphics[width=0.47\textwidth]{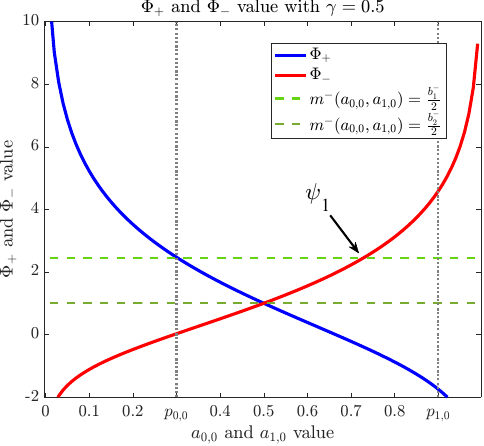}
    \caption{Illustration of the uniqueness of $S_{b}^-$ in appendix \ref{sec:appendix_sbnonempty}}
    \label{fig:psi_proof}
\end{figure}

Let us first consider the two functions $\Phi_+(a_{0,0})$ and 
$\Phi_-(a_{1,0})$, as illustrated in Fig.~\ref{fig:psi_proof}. 
The two functions meet at the point 
$a_{0,0}=a_{1,0}=\frac{1}{2}$. 
Their intersection with the darker green dashed line corresponds to
$(a_{0,0},a_{1,0})=(\frac{1}{2},\frac{1}{2})$, which is the only
element in $S^-_{b_2^{-}}$. For the case shown in Fig.~\ref{fig:psi_proof},
their intersections with the lighter green dashed line correspond to
$(a_{0,0},a_{1,0})=(p_{0,0},\psi_1)$, which is the only element in
$S^-_{b_1^{-}}$. For the other case,
$|\frac{1}{2}-p_{0,0}|>|\frac{1}{2}-p_{1,0}|$, the corresponding
endpoint is $(\psi_0,p_{1,0})$.

The figure suggests that, for every $b\in[b_1^{-},b_2^{-}]$, there exists a
unique pair $(a_{0,0},a_{1,0})$ satisfying
\[
\Phi_+(a_{0,0})=\Phi_-(a_{1,0}),
\qquad
m^{-}(a_{0,0},a_{1,0})=\frac{b}{2}.
\]
To establish this result, we prove the following two claims.

\begin{enumerate}
    \item \textit{Claim 1: $\Phi_+$ is strictly decreasing on
    $(0,\frac{1}{2}]$, while $\Phi_-$ is strictly increasing on
    $[\frac{1}{2},1)$.}

    We prove this claim by computing the derivatives:
    \begin{align}
        \frac{\partial \Phi_{+}}{\partial a}
        &=
        \frac{-(1-2\gamma+\gamma p_{0,0}+\gamma p_{1,0})}{1-a}
        -\frac{1+\gamma p_{0,0}+\gamma p_{1,0}}{a}
        \nonumber\\
        &=
        \frac{2\gamma a-1-\gamma p_{0,0}-\gamma p_{1,0}}
        {(1-a)a}
        <0,
        \qquad
        \forall a\in(0,\frac{1}{2}],
        \label{eq:phi0_decreasing}
        \\
        \frac{\partial \Phi_{-}}{\partial a}
        &=
        \frac{-(-1-2\gamma+\gamma p_{0,0}+\gamma p_{1,0})}{1-a}
        -\frac{-1+\gamma p_{0,0}+\gamma p_{1,0}}{a}
        \nonumber\\
        &=
        \frac{2\gamma a+1-\gamma p_{0,0}-\gamma p_{1,0}}
        {(1-a)a}
        >0,
        \qquad
        \forall a\in [\frac{1}{2},1).
        \label{eq:phi1_increasing}
    \end{align}
    Therefore, $\Phi_+$ is strictly decreasing and $\Phi_-$ is
    strictly increasing over the corresponding intervals. Moreover,
    as mentioned above, the two functions meet at $a=\frac{1}{2}$,
    where
    \[
        \Phi_+(\tfrac{1}{2})
        =
        \Phi_-(\tfrac{1}{2})
        =
        -2\gamma\log\tfrac{1}{2}.
    \]

    The endpoint values above also guarantee the existence of a pair
$(a_{0,0},a_{1,0})$ satisfying
\[
    \Phi_+(a_{0,0})=\Phi_-(a_{1,0})
\]

Since $\Phi_+$ is continuous and strictly decreasing on
$[p_{0,0},\frac{1}{2}]$, and $\Phi_-$ is continuous and strictly
increasing on $[\frac{1}{2},\psi_1]$, the two functions cover the
same range of values over these intervals. Therefore, for every
$a_{0,0}\in[p_{0,0},\frac{1}{2}]$, there exists a unique
$a_{1,0}\in[\frac{1}{2},\psi_1]$ such that
\[
    \Phi_+(a_{0,0})=\Phi_-(a_{1,0}).
\]
Hence, along this equality, $a_{1,0}$ can be written as a function of
$a_{0,0}$, denoted by $a_{1,0}=f(a_{0,0})$. 
Substituting this into the equality gives
\[
    \Phi_+(a_{0,0})
    =
    \Phi_-\bigl(f(a_{0,0})\bigr).
\]
Differentiating both sides with respect to $a_{0,0}$ and applying the
chain rule, we obtain
\[
    \Phi_+'(a_{0,0})
    =
    \Phi_-'(a_{1,0})
    \frac{d a_{1,0}}{d a_{0,0}},
\]
and therefore
\begin{equation}
    \frac{d a_{1,0}}{d a_{0,0}}
    =
    \frac{\Phi_+'(a_{0,0})}
         {\Phi_-'(a_{1,0})}
    <0.
    \label{eq:a1_a0_monotonic}
\end{equation}
Thus, $a_{1,0}$ strictly decreases as $a_{0,0}$ increases along the
curve $\Phi_+(a_{0,0})=\Phi_-(a_{1,0})$.

    \item \textit{Claim 2: Along the curve
    $\Phi_+(a_{0,0})=\Phi_-(a_{1,0})$, the function
    $m^{-}(a_{0,0},a_{1,0})$ is strictly increasing with respect to
    $a_{0,0}$.}

    Recall that
    \[
        d_0
        =
        \frac{\mathbf d_0^\gamma+\gamma a_{1,0}}
        {1-\gamma a_{0,0}+\gamma a_{1,0}},
        \qquad
        d_1
        =
        \frac{1-\gamma a_{0,0}-\mathbf d_0^\gamma}
        {1-\gamma a_{0,0}+\gamma a_{1,0}},
    \]
    and
    \[
        m^{-}(a_{0,0},a_{1,0})
        =
        d_0(a_{0,0}-p_{0,0})
        +
        d_1(p_{1,0}-a_{1,0}).
    \]
    Direct differentiation gives
    \begin{align}
        \frac{\partial m^{-}}{\partial a_{0,0}}
        &=
        \frac{
        (\mathbf d_0^\gamma+\gamma a_{1,0})
        \left(
        1+2\gamma a_{1,0}
        -\gamma p_{0,0}
        -\gamma p_{1,0}
        \right)}
        {(1-\gamma a_{0,0}+\gamma a_{1,0})^2}
        >0,
        \label{eq:m_partial_a0}
        \\
        \frac{\partial m^{-}}{\partial a_{1,0}}
        &=
        -\frac{
        (1-\gamma a_{0,0}-\mathbf d_0^\gamma)
        \left(
        1-2\gamma a_{0,0}
        +\gamma p_{0,0}
        +\gamma p_{1,0}
        \right)}
        {(1-\gamma a_{0,0}+\gamma a_{1,0})^2}
        <0.
        \label{eq:m_partial_a1}
    \end{align}
    Combining \eqref{eq:a1_a0_monotonic},
    \eqref{eq:m_partial_a0}, and \eqref{eq:m_partial_a1}, we obtain
    \begin{align}
        \frac{d}{d a_{0,0}}
        m^{-}(a_{0,0},f(a_{0,0}))
        &=
        \frac{\partial m^{-}}{\partial a_{0,0}}
        +
        \frac{\partial m^{-}}{\partial a_{1,0}}
        \frac{d a_{1,0}}{d a_{0,0}}
        >0.
        \label{eq:m_curve_monotonic}
    \end{align}
    Therefore, $m^{-}(a_{0,0},a_{1,0})$ is strictly increasing along
    the curve $\Phi_+(a_{0,0})=\Phi_-(a_{1,0})$.

    It remains to identify the values of $m^{-}$ at the two endpoints.
    For the lower boundary $b=b_1^{-}$, we consider the two possible
    cases. If
    $|\frac{1}{2}-p_{1,0}|
    \geq
    |\frac{1}{2}-p_{0,0}|$, then the endpoint is
    $(p_{0,0},\psi_1)$, and
    \begin{align}
        &d_0(p_{0,0}-p_{0,0})
        +d_1(p_{1,0}-\psi_1)
        =d_1(p_{1,0}-\psi_1)
        \nonumber\\
        &\quad=
        \left(
        \frac{
        1-\gamma p_{0,0}-\mathbf d_0^\gamma}
        {1-\gamma p_{0,0}+\gamma\psi_1}
        \right)
        (p_{1,0}-\psi_1)
        =
        \frac{b_1^{-}}{2}.
        \label{eq:p0_psi1_sol2bl}
    \end{align}
    If
    $|\frac{1}{2}-p_{0,0}|
    >
    |\frac{1}{2}-p_{1,0}|$, then the endpoint is
    $(\psi_0,p_{1,0})$, and
    \begin{align}
        &d_0(\psi_0-p_{0,0})
        +d_1(p_{1,0}-p_{1,0})
        =d_0(\psi_0-p_{0,0})
        \nonumber\\
        &\quad=
        \left(
        \frac{
        \mathbf d_0^\gamma+\gamma p_{1,0}}
        {1-\gamma\psi_0+\gamma p_{1,0}}
        \right)
        (\psi_0-p_{0,0})
        =
        \frac{b_1^{-}}{2}.
        \label{eq:psi0_p1_sol2bl}
    \end{align}

    For the upper boundary $b=b_2^{-}$, the endpoint is
    $(\frac{1}{2},\frac{1}{2})$. Substituting this point into $m$
    gives
    \begin{align}
        &d_0\left(\frac{1}{2}-p_{0,0}\right)
        +d_1\left(p_{1,0}-\frac{1}{2}\right)
        =
        \frac{1}{2}(d_0-d_1)
        -d_0p_{0,0}
        +d_1p_{1,0}
        \nonumber\\
        &=
        \frac{1}{2}
        \left(
        \frac{
        \mathbf d_0^\gamma+\frac{\gamma}{2}
        -1+\frac{\gamma}{2}+\mathbf d_0^\gamma}
        {1-\frac{\gamma}{2}+\frac{\gamma}{2}}
        \right)
        -p_{0,0}
        \left(
        \frac{\mathbf d_0^\gamma+\frac{\gamma}{2}}
        {1-\frac{\gamma}{2}+\frac{\gamma}{2}}
        \right)
        +
        p_{1,0}
        \left(
        \frac{
        1-\frac{\gamma}{2}-\mathbf d_0^\gamma}
        {1-\frac{\gamma}{2}+\frac{\gamma}{2}}
        \right)
        \nonumber\\
        &=
        \left(
        \mathbf d_0^\gamma+\frac{\gamma}{2}
        \right)
        (1-p_{0,0}-p_{1,0})
        +p_{1,0}-\frac{1}{2}
        =
        \frac{b_2^{-}}{2}.
        \label{eq:0.5_is_sol2bh}
    \end{align}
\end{enumerate}

Therefore, along the curve
$\Phi_+(a_{0,0})=\Phi_-(a_{1,0})$, the continuous function
$m^{-}(a_{0,0},a_{1,0})$ increases strictly from
$\frac{b_1^{-}}{2}$ to $\frac{b_2^{-}}{2}$. Consequently, for every
$b\in[b_1^{-},b_2^{-}]$, there exists exactly one pair
\[
(a_{0,0},a_{1,0})
\in
[p_{0,0},\frac{1}{2}]
\times
[\frac{1}{2},p_{1,0}]
\]
satisfying
\[
\Phi_+(a_{0,0})=\Phi_-(a_{1,0}),
\qquad
m^{-}(a_{0,0},a_{1,0})=\frac{b}{2}.
\]
Hence, $S_b^{-}$ is a singleton set for every
$b\in[b_1^{-},b_2^{-}]$.

\section{Proof of Reduction of Cases} \label{pf:cor1}
In scenarios where $p_{0,0}, p_{1,0} \leq \frac{1}{2}$, we interchange state $0 $ to $1^{\prime}$ and state $1 $ to $0^{\prime}$ as the order of states is inconsequential. We first note that in our setting, $p_{0,0} = p_{0\rightarrow 0}$ and $p_{1,0} = p_{1\rightarrow 0}$, meaning that $p_{0,0}$ is the probability from state 0 to 0 and $p_{1,0}$ is the probability from state 1 to 0. After renaming the state from $0, 1$ to $1^{\prime}, 0^{\prime}$, the probability $p_{0,0}, p_{1,0}$ now becomes 
\begin{align}
    & p_{0,0} = p_{0 \rightarrow 0} \overset{\tiny{\text{rename}}}{=} p_{1^{\prime} \rightarrow 1^{\prime}} \triangleq 1-p^{\prime}_{1,0} \label{eq:state_rename0}\\
    & p_{1,0} = p_{1 \rightarrow 0} \overset{\tiny{\text{rename}}}{=} p_{0^{\prime} \rightarrow 1^{\prime}} \triangleq 1-p^{\prime}_{0,0}.\label{eq:state_rename1}
\end{align}
Additionally, the initial state distributions are swapped, so $d_{0} = d_1'$ and $d_{1} = d_0'$.
Therefore, for case $p_{0,0}, p_{1,0} \leq \frac{1}{2}$, the renamed optimization problem will have $p^{\prime}_{0,0},p^{\prime}_{1,0} \geq \frac{1}{2}$ with initial state distribution $d_0',d_1'$. Use the results in Theorem \ref{thm:p0p1sameside} to solve for optimal $a^{\prime}_{0,0}, a^{\prime}_{1,0}$, then convert back to $a_{0,0}, a_{1,0}$.
\begin{align}
    a^{\prime}_{0,0} = a_{0^{\prime} \rightarrow 0^{\prime}} \overset{\tiny{\text{recover}}}{=} a_{1 \rightarrow 1} = 1 - a_{1,0} & \Rightarrow a_{1,0} = 1 - a^{\prime}_{0,0} \\
    a^{\prime}_{1,0} = a_{1^{\prime} \rightarrow 0^{\prime}} \overset{\tiny{\text{recover}}}{=} a_{0 \rightarrow 1} = 1 - a_{0,0} & \Rightarrow a_{0,0} = 1 - a^{\prime}_{1,0}
\end{align}

\section{Proof of Theorem \ref{thm:4s_regimeI}} \label{sec:4s_regimeI_pf}

To proof this theorem, we need to solve the following optimization problem which is equivalent to problem $\mathbf{P3\text{-}(2,2)}$:
\begin{align}
    \max_{\substack{
    (d_{\rvs,0},x_{\rvs,0},c_{\rvs,0})_{\rvs\in\mathsf S}
    }}
    & \; \; \sum_{\rvs \in \mathsf{S}} d_{\rvs,0} H\!\left(\tfrac{x_{\rvs,0}}{d_{\rvs,0}}\right) 
    \label{eq:x0_obj_func_TV_4s_with_c} \\
    \text{subject to:} & ~ d_{00,0} =\rvd_{00}^{\gamma}+ \gamma( x_{00,0} + x_{01,0}) \label{eq:x0_constraint1_TV_4s_with_c}\\
    & ~ d_{01,0} =\rvd_{01}^{\gamma}+ \gamma( x_{10,0} + x_{11,0}) \label{eq:x0_constraint2_TV_4s_with_c}\\
    & ~ d_{10,0} =\rvd_{10}^{\gamma}+ \gamma ( d_{00,0}- x_{00,0}) + \gamma ( d_{01,0}- x_{01,0})
    \label{eq:x0_constraint3_TV_4s_with_c} \\
    & ~ d_{11,0} =\rvd_{11}^{\gamma}+ \gamma ( d_{10,0}- x_{10,0}) + \gamma ( d_{11,0}- x_{11,0})
    \label{eq:x0_constraint4_TV_4s_with_c} \\
    & ~ -c_{\rvs,0} \leq x_{\rvs,0}-d_{\rvs,0}p_{\rvs,0} \leq c_{\rvs,0}, \; \forall \rvs\in\mathsf S
    \label{eq:x0_constraint5_TV_4s_with_c}\\
    & ~ \sum_{\rvs \in \mathsf{S}} c_{\rvs,0}\leq \frac{b}{2}
    \label{eq:x0_constraint6_TV_4s_with_c}\\
    & ~ 0 \leq x_{\rvs,0} \leq d_{\rvs,0}, \; c_{\rvs,0}\geq0, \; \forall \rvs \in \mathsf{S}.
    \label{eq:x0_constraint7_TV_4s_with_c}
\end{align}
We first provide the optimal primal variable solution to this problem in regime I:
\begin{align}
d_{\rvs,0}(b)
&=
\rvd^{\mathrm{base}}_{\rvs}
+
\frac{\gamma\sigma_{s_1}b}{2}
\left(
[\mM(0)]_{\rvs,n_1(s_1)}
-
[\mM(0)]_{\rvs,n_0(s_1)}
\right),
\qquad \rvs\in\mathsf S,
\\
x_{\rvs,0}(b)
&=
\begin{cases}
d_{s_1, 0}(b)p_{s_1, 0}
-\sigma_{s_1}\dfrac{b}{2},
&
\rvs=s_1,
\\[2mm]
d_{\rvs,0}(b)p_{\rvs,0},
&
\text{otherwise},
\end{cases}
\\
c_{\rvs,0}(b)
&=
\begin{cases}
\dfrac{b}{2},
&
\rvs=s_1,
\\
0,
&
\text{otherwise}.
\end{cases}
\end{align}
The $x_{\rvs,0}$ expressions follow from
\begin{equation*}
x_{\rvs,0}(b)=d_{\rvs,0}(b)a_{\rvs,0}(b),
\end{equation*}
together with the active TV-budget equality
\begin{equation*}
d_{s_1, 0}(b)u_{s_1}(b)=\frac{b}{2},
\end{equation*}
where \(a_{\rvs,0}(b)\) is the Regime-I optimal policy given in
Theorem~\ref{thm:4s_regimeI}:
\begin{equation*}
a_{\rvs,0}
=
\begin{cases}
p_{\rvs,0}-\sigma_{\rvs}u_{s_1}(b),
&
\rvs=s_1,
\\
p_{\rvs,0},
&
\text{otherwise}.
\end{cases}
\end{equation*}
In particular, for the active state \(s_1\),
\begin{equation*}
d_{s_1, 0}(b)
=
\rvd^{\mathrm{base}}_{s_1}
+
\frac{\gamma\sigma_{s_1}b}{2}
\Delta\mM(0)_{s_1}.
\end{equation*}

\begin{enumerate}
\item Primal feasibility

The constraints \eqref{eq:x0_constraint1_TV_4s_with_c}--\eqref{eq:x0_constraint4_TV_4s_with_c} are obtained from the original relation
\begin{align}
    \rvd &= \rvd^\gamma + \gamma\mA(b)^\top\rvd
    \Rightarrow
    \rvd = \left(I-\gamma \mA(b)^\top \right)^{-1} \rvd^\gamma = \mM(b)\rvd^\gamma .
\end{align}
In regime I, only the state \(s_1\) changes its action from $p_{s_1, 0}$ by $-\sigma_{s_1}u_{s_1}$. Consequently, the action matrix can be written as
\[
    \mA(b)
    =
    \mP
    +
    \sigma_{s_1}u_{s_1}(b)
    e_{s_1}
    \left(
    e_{n_1(s_1)}-e_{n_0(s_1)}
    \right)^\top,
\]
where \(e_j\in\mathbb R^4\) is the \(j\)-th standard basis vector. Using the Sherman--Morrison formula, we obtain
\begin{align}
    \mM(b)
    &=
    \mM(0)
    +
    \frac{
        \gamma\sigma_{s_1}u_{s_1}
        \mM(0)
        \left(e_{n_1(s_1)}-e_{n_0(s_1)}\right)
        e_{s_1}^{\top}\mM(0)
    }{
        1-\gamma\sigma_{s_1}u_{s_1}
        e_{s_1}^{\top}\mM(0)
        \left(e_{n_1(s_1)}-e_{n_0(s_1)}\right)
    }, \label{eq:M(b)_SM_formula}
\end{align}
and thus
\begin{align}
    \rvd
    &=
    \mM(b)\rvd^\gamma
    =
    \rvd^{\mathrm{base}}
    +
    \frac{
        \gamma\sigma_{s_1}u_{s_1}
        \mM(0)
        \left(e_{n_1(s_1)}-e_{n_0(s_1)}\right)
        e_{s_1}^{\top}\rvd^{\mathrm{base}}
    }{
        1-\gamma\sigma_{s_1}u_{s_1}\Delta\mM(0)_{s_1}
    }
    \\
    \Rightarrow
    d_{\rvs,0}
    &=
    \rvd^{\mathrm{base}}_{\rvs}
    +
    \frac{
        \gamma\sigma_{s_1}u_{s_1}\rvd^{\mathrm{base}}_{s_1}
    }{
        1-\gamma\sigma_{s_1}u_{s_1}\Delta\mM(0)_{s_1}
    }
    \left(
        [\mM(0)]_{\rvs,n_1(s_1)}
        -
        [\mM(0)]_{\rvs,n_0(s_1)}
    \right).
\end{align}
Substituting \(u_{s_1}(b)\) from Theorem~\ref{thm:4s_regimeI} yields
\begin{align*}
d_{\rvs,0}(b)
&=
\rvd^{\mathrm{base}}_{\rvs}
+
\frac{\gamma\sigma_{s_1}b}{2}
\left(
[\mM(0)]_{\rvs,n_1(s_1)}
-
[\mM(0)]_{\rvs,n_0(s_1)}
\right),
\qquad \rvs\in\mathsf S.
\end{align*}
In particular,
\begin{align*}
d_{s_1, 0}(b)u_{s_1}(b)
&=
\left(
\rvd^{\mathrm{base}}_{s_1}
+
\frac{\gamma\sigma_{s_1}b}{2}
\Delta\mM(0)_{s_1}
\right)
\frac{b}{
2\rvd^{\mathrm{base}}_{s_1}
+b\gamma\sigma_{s_1}\Delta\mM(0)_{s_1}}
=
\frac{b}{2}.
\end{align*}
Finally, since \(x_{\rvs,0}\) is defined via \(x_{\rvs,0}=d_{\rvs,0}a_{\rvs,0}\), this implies that constraints \eqref{eq:x0_constraint1_TV_4s_with_c}--\eqref{eq:x0_constraint4_TV_4s_with_c} are fulfilled.

For constraints \eqref{eq:x0_constraint5_TV_4s_with_c} and \eqref{eq:x0_constraint6_TV_4s_with_c}, these two conditions are clearly satisfied by directly substituting the solutions for \(d_{\rvs,0}\), \(x_{\rvs,0}\), and \(c_{\rvs,0}\). In particular, at state $s_1$,
\begin{equation*}
\left|
x_{s_1, 0}-d_{s_1, 0}p_{s_1, 0}
\right|
=
d_{s_1, 0}u_{s_1}(b)
=
\frac{b}{2},
\end{equation*}
while this quantity is zero for every $\rvs\neq s_1$. For the final constraint \eqref{eq:x0_constraint7_TV_4s_with_c}, since \(x_{\rvs,0}=d_{\rvs,0}a_{\rvs,0}\) and \(0<a_{\rvs,0}<1\) throughout Regime I, we obtain \(0<x_{\rvs,0}<d_{\rvs,0}\).

\item Dual variables and dual feasibility

The following dual variables are associated with the constraints of
this problem:
\begin{align*}
\lambda(b)\geq0
&:
\quad
\sum_{\rvs\in\mathsf S}c_{\rvs,0}-\frac{b}{2}\leq0,
\\
\alpha_{\rvs}\geq0
&:
\quad
x_{\rvs,0}-d_{\rvs,0}p_{\rvs,0}-c_{\rvs,0}\leq0,
\\
\beta_{\rvs}\geq0
&:
\quad
d_{\rvs,0}p_{\rvs,0}-x_{\rvs,0}-c_{\rvs,0}\leq0,
\\
\omega_{\rvs}\in\mathbb R
&:
\quad
\text{the flow-conservation equality associated with state }\rvs,
\\
\nu_{\rvs}\geq0
&:
\quad
-x_{\rvs,0}\leq0,
\\
\mu_{\rvs}\geq0
&:
\quad
x_{\rvs,0}-d_{\rvs,0}\leq0,
\\
\rho_{\rvs}\geq0
&:
\quad
-c_{\rvs,0}\leq0.
\end{align*}

For the proposed Regime-I solution, set the dual variable $\lambda$ to
\begin{align}
\lambda
&=
\lambda(b)=\frac{\sigma_{s_1}K^0_{s_1}(\mA(b))}{1-\sigma_{s_1}\gamma \left([\mM(b)^\top\mD(\mA(b),\mP)]_{n_0(s_1)} - [\mM(b)^\top\mD(\mA(b),\mP)]_{n_1(s_1)}\right)} \label{eq:lambda_def_(2,2)}
\end{align}
which follows from solving the equality
\begin{align}
    \lambda(b) = \sigma_{s_1}K_{s_1}^{\lambda(b)}
\bigl(\mA(b)\bigr).
\end{align}
The denominator in \eqref{eq:lambda_def_(2,2)} is strictly positive; this can be checked by substituting the expression for $\mA(b)$ and then using \eqref{eq:M(b)_SM_formula} for $\mM(b)$.
In addition, we choose the remaining dual variables by
\begin{align}
\alpha_{\rvs}
&=
\max\left\{-K_{\rvs}^{\lambda(b)}\bigl(\mA(b)\bigr),0\right\}, ~
\beta_{\rvs}
=
\max\left\{K_{\rvs}^{\lambda(b)}\bigl(\mA(b)\bigr),0\right\}, ~
\rho_{\rvs}
=
\lambda(b)
-
\left|
K_{\rvs}^{\lambda(b)}\bigl(\mA(b)\bigr)
\right|,
\\
\omega_{\rvs}
&=
W_{\rvs}^{\lambda(b)}\bigl(\mA(b)\bigr), ~
\nu_{\rvs}
=
\mu_{\rvs}
=
0.
\end{align}
From the above, it follows that $\alpha_{\rvs}, \nu_\rvs, \mu_\rvs$ and $\beta_{\rvs}$ are nonnegative. We next show that $\lambda$ and $\rho_{\rvs}$, for all $\rvs$, are likewise nonnegative. To do so, we first prove the following statement.

\underline{\textbf{Claim:}
$\sign\left(K_{s_1}^{\lambda(b)}(\mA(b))\right)=\sigma_{s_1}$ in Regime I.} \\

\begin{proof}
In Theorem~\ref{thm:4s_regimeI}, we assume the condition
\begin{equation*}
\sigma_{s_1}K_{s_1}^{0}(\mP)>0,
\end{equation*}
which implies that $K_{s_1}^{0}(\mP)$ must have the same sign as $\sigma_{s_1}$.
Since \(\mA(b)\to\mP\) as \(b\to0^+\), we have \(\mD(\mA(b),\mP)\to0\). Hence \(K^{\lambda(b)}_{\rvs}(\mA(b)) \to K^0_{\rvs}(\mP)\) as \(b\to0^+\). Therefore, \(K^{\lambda(b)}_{\rvs}(\mP)\) is continuous in \(b\) on \([0,b_1)\). By the continuity of \(K_{\rvs}^{\lambda}(\mA)\), we obtain
\[
\sigma_{s_1}
K_{s_1}^{\lambda(b)}
\bigl(\mA(b)\bigr)
>0 \Rightarrow \sigma_{s_1} = \sign(K_{s_1}^{\lambda(b)}
(\mA(b)))
\]
for all sufficiently small \(b>0\).

Suppose that the sign of
\(K_{s_1}^{\lambda(b)}(\mA(b))\) changes before the end of Regime I.
By continuity, there must exist some \(b'\in(0,b_1)\) such that
\begin{equation*}
K_{s_1}^{\lambda(b')}
\bigl(\mA(b')\bigr)
=
0.
\end{equation*}
This implies
\begin{equation*}
I_{s_1}^{\lambda(b')}
\bigl(\mA(b')\bigr)
=
0.
\end{equation*}
However, by the definition of the Regime-I boundary \(b_1\),
\begin{equation*}
I_{\rvs}^{\lambda(b)}
\bigl(\mA(b)\bigr)
<
I_{s_1}^{\lambda(b)}
\bigl(\mA(b)\bigr),
\qquad
\rvs\neq s_1,\quad b\in[0,b_1).
\end{equation*}
Therefore, at \(b=b'\), we would have
\begin{equation*}
I_{\rvs}^{\lambda(b')}
\bigl(\mA(b')\bigr)
<
0,
\qquad
\rvs\neq s_1,
\end{equation*}
which is impossible since
\(I_{\rvs}^{\lambda}(\mA)=|K_{\rvs}^{\lambda}(\mA)|\geq0\).
Therefore, the sign of
\(K_{s_1}^{\lambda(b)}(\mA(b))\) remains unchanged throughout Regime I,
and thus the claim is proved.
\end{proof}

By the definition of $\lambda$ in \eqref{eq:lambda_def_(2,2)}, we have $
K_{s_1}^{\lambda(b)}
\bigl(\mA(b)\bigr)
=
\sigma_{s_1}\lambda(b)$.
Because $\sigma_{s_1}$ shares the same sign as $K_{s_1}^{\lambda(b)}
\bigl(\mA(b)\bigr)$ throughout regime I, it follows that
\begin{equation*}
\lambda = \lambda(b) = \bigl|K_{s_1}^{\lambda(b)}
\bigl(\mA(b)\bigr)\bigr|.
\end{equation*}
Consequently, $\lambda$ is nonnegative and $\rho_{s_1}=0$ for all of regime I.
In contrast, for any inactive state \(\rvs\neq s_1\), we obtain
\begin{equation*}
\left|
K_{\rvs}^{\lambda(b)}
\bigl(\mA(b)\bigr)
\right|
=
I_{\rvs}^{\lambda(b)}
\bigl(\mA(b)\bigr)
<
I_{s_1}^{\lambda(b)}
\bigl(\mA(b)\bigr)
=
\lambda(b),
\end{equation*}
where the strict inequality is implied by the definition of \(b_1\). Hence,
\begin{equation*}
\rho_\rvs = \lambda(b) - \left| K_{\rvs}^{\lambda(b)}
\bigl(\mA(b)\bigr)
\right|
\geq 0
,
\qquad
\forall \rvs\in\mathsf S.
\end{equation*}

\item Complementary slackness

We next verify that the proposed primal and dual variables satisfy the
complementary-slackness conditions. The conditions corresponding to the
inequality constraints are
\begin{align*}
\lambda(b)
\left(
\sum_{\rvs\in\mathsf S}c_{\rvs,0}-\frac{b}{2}
\right)
&=0,
\\
\alpha_{\rvs}
\left(
x_{\rvs,0}-d_{\rvs,0}p_{\rvs,0}-c_{\rvs,0}
\right)
&=0,
\\
\beta_{\rvs}
\left(
d_{\rvs,0}p_{\rvs,0}-x_{\rvs,0}-c_{\rvs,0}
\right)
&=0,
\\
\nu_{\rvs}x_{\rvs,0}
&=0,
\\
\mu_{\rvs}
\left(
x_{\rvs,0}-d_{\rvs,0}
\right)
&=0,
\\
\rho_{\rvs}c_{\rvs,0}
&=0.
\end{align*}

We verify these conditions separately. Since
$\sum_{\rvs\in\mathsf S}c_{\rvs,0}=b/2$, thus
\begin{equation*}
\lambda(b)
\left(
\sum_{\rvs\in\mathsf S}c_{\rvs,0}-\frac{b}{2}
\right)
=
0.
\end{equation*}

Next, consider the active state \(s_1\). Using the Regime-I solution together with \(c_{s_1, 0}=b/2\), we obtain
\begin{equation*}
x_{s_1, 0}
-
d_{s_1, 0}p_{s_1, 0}
=
-\sigma_{s_1}\frac{b}{2}.
\end{equation*}
If $\sigma_{s_1}=1$, then
\begin{align*}
x_{s_1, 0}
-d_{s_1, 0}p_{s_1, 0}
-c_{s_1, 0}
&=
-b<0,
\\
d_{s_1, 0}p_{s_1, 0}
-x_{s_1, 0}
-c_{s_1, 0}
&=
0.
\end{align*}
Combining this with the previously derived result
\begin{equation*}
K_{s_1}^{\lambda(b)}
\bigl(\mA(b)\bigr)
=
\lambda(b),
\end{equation*}
we obtain
\begin{equation*}
\alpha_{s_1}=0,
\qquad
\beta_{s_1}=\lambda(b).
\end{equation*}
Thus, the two complementary-slackness conditions corresponding to the active state are fulfilled. When $\sigma_{s_1}=-1$, the first inequality is active instead, and
\begin{equation*}
\alpha_{s_1}=\lambda(b),
\qquad
\beta_{s_1}=0,
\end{equation*}
so the same conclusion follows.

For every inactive state \(\rvs\neq s_1\), the action remains unchanged from the base policy. Hence,
\begin{equation*}
x_{\rvs,0}
=
d_{\rvs,0}p_{\rvs,0},
\qquad
c_{\rvs,0}=0.
\end{equation*}
It follows that
\begin{equation*}
x_{\rvs,0}
-d_{\rvs,0}p_{\rvs,0}
-c_{\rvs,0}
=
d_{\rvs,0}p_{\rvs,0}
-x_{\rvs,0}
-c_{\rvs,0}
=
0,
\end{equation*}
and therefore the complementary-slackness conditions associated with
\(\alpha_{\rvs}\) and \(\beta_{\rvs}\) hold regardless of their values.

It remains to verify the complementary-slackness condition associated
with \(\rho_{\rvs}\). For the active state $s_1$, we have
\begin{equation*}
\left|
K_{s_1}^{\lambda(b)}
\bigl(\mA(b)\bigr)
\right|
=
\lambda(b)
\Rightarrow
\rho_{s_1}=0
\Rightarrow
\rho_{s_1}c_{s_1, 0}=0.
\end{equation*}
For all other states, \(c_{\rvs,0}=0\), and hence
\begin{equation*}
\rho_{\rvs}c_{\rvs,0}=0.
\end{equation*}

Finally, throughout Regime I, $0<x_{\rvs,0}<d_{\rvs,0}$,
so the constraints \(-x_{\rvs,0}\leq0\) and
\(x_{\rvs,0}-d_{\rvs,0}\leq0\) are inactive, thus we choose
\begin{equation*}
\nu_{\rvs}=\mu_{\rvs}=0,
\end{equation*}
and the complementary-slackness conditions are satisfied.

\item Stationarity

The Lagrangian of this problem is given by
\begin{align*}
L
=&
-\sum_{\rvs\in\mathsf S}
d_{\rvs,0}
H\!\left(
\frac{x_{\rvs,0}}{d_{\rvs,0}}
\right)
+
\lambda(b)
\left(
\sum_{\rvs\in\mathsf S}c_{\rvs,0}
-\frac{b}{2}
\right)
+
\sum_{\rvs\in\mathsf S}
\alpha_{\rvs}
\left(
x_{\rvs,0}
-d_{\rvs,0}p_{\rvs,0}
-c_{\rvs,0}
\right)
\\
&+
\sum_{\rvs\in\mathsf S}
\beta_{\rvs}
\left(
d_{\rvs,0}p_{\rvs,0}
-x_{\rvs,0}
-c_{\rvs,0}
\right)
+
\omega_{00}
\left(
d_{00,0}-\rvd_{00}^{\gamma}
-\gamma x_{00,0}
-\gamma x_{01,0}
\right)
\\
&+
\omega_{01}
\left(
d_{01,0}-\rvd_{01}^{\gamma}
-\gamma x_{10,0}
-\gamma x_{11,0}
\right)
+
\omega_{10}
\left(
d_{10,0}-\rvd_{10}^{\gamma}
-\gamma(d_{00,0}-x_{00,0})
-\gamma(d_{01,0}-x_{01,0})
\right)
\\
&+
\omega_{11}
\left(
d_{11,0}-\rvd_{11}^{\gamma}
-\gamma(d_{10,0}-x_{10,0})
-\gamma(d_{11,0}-x_{11,0})
\right)
-
\sum_{\rvs\in\mathsf S}\nu_{\rvs}x_{\rvs,0}
\\
&+
\sum_{\rvs\in\mathsf S}
\mu_{\rvs}
\left(
x_{\rvs,0}-d_{\rvs,0}
\right)
-
\sum_{\rvs\in\mathsf S}
\rho_{\rvs}c_{\rvs,0}.
\end{align*}

Using \(a_{\rvs,0}=x_{\rvs,0}/d_{\rvs,0}\), the stationarity conditions with respect to \(x_{\rvs,0}\), \(d_{\rvs,0}\), and \(c_{\rvs,0}\) can be written as
\begin{align*}
\frac{\partial L}{\partial x_{\rvs,0}}
&=
\log\left(\frac{a_{\rvs,0}}{1-a_{\rvs,0}}\right)
+\alpha_{\rvs}
-\beta_{\rvs}
-\gamma\omega_{n_0(\rvs)}
+\gamma\omega_{n_1(\rvs)}
-\nu_{\rvs}
+\mu_{\rvs}
=0,
\\
\frac{\partial L}{\partial d_{\rvs,0}}
&=
\log(1-a_{\rvs,0})
+\omega_{\rvs}
-\gamma\omega_{n_1(\rvs)}
+p_{\rvs,0}
\left(
\beta_{\rvs}-\alpha_{\rvs}
\right)
-\mu_{\rvs}
=0,
\\
\frac{\partial L}{\partial c_{\rvs,0}}
&=
\lambda(b)
-\alpha_{\rvs}
-\beta_{\rvs}
-\rho_{\rvs}
=0.
\end{align*}

We first consider the stationarity condition with respect to
\(x_{\rvs,0}\). From the definitions of the dual variables,
\begin{equation*}
\beta_{\rvs}-\alpha_{\rvs}
=
K_{\rvs}^{\lambda(b)}
\bigl(\mA(b)\bigr),
\qquad
\omega_{\rvs}
=
W_{\rvs}^{\lambda(b)}
\bigl(\mA(b)\bigr),
\end{equation*}
and \(\nu_{\rvs}=\mu_{\rvs}=0\). Therefore,
\begin{align*}
\frac{\partial L}{\partial x_{\rvs,0}}
&=
\log\left(\frac{a_{\rvs,0}}{1-a_{\rvs,0}}\right)
-\gamma
W_{n_0(\rvs)}^{\lambda(b)}
\bigl(\mA(b)\bigr)
+\gamma
W_{n_1(\rvs)}^{\lambda(b)}
\bigl(\mA(b)\bigr)
-
K_{\rvs}^{\lambda(b)}
\bigl(\mA(b)\bigr)
=
0,
\end{align*}
where the last equality follows directly from the definition
\begin{equation*}
K_{\rvs}^{\lambda}(\mA)
=
\log\left(\frac{a_{\rvs,0}}{1-a_{\rvs,0}}\right)
-\gamma
W_{n_0(\rvs)}^{\lambda}(\mA)
+\gamma
W_{n_1(\rvs)}^{\lambda}(\mA).
\end{equation*}

For \(c_{\rvs,0}\), using
\begin{equation*}
\alpha_{\rvs}+\beta_{\rvs}
=
\left|
K_{\rvs}^{\lambda(b)}
\bigl(\mA(b)\bigr)
\right| \text{ and }
\rho_{\rvs}
=
\lambda(b)
-
\left|
K_{\rvs}^{\lambda(b)}
\bigl(\mA(b)\bigr)
\right|,
\end{equation*}
we obtain
\begin{equation*}
\frac{\partial L}{\partial c_{\rvs,0}}
=
\lambda(b)
-
\left|
K_{\rvs}^{\lambda(b)}
\bigl(\mA(b)\bigr)
\right|
-
\left(
\lambda(b)
-
\left|
K_{\rvs}^{\lambda(b)}
\bigl(\mA(b)\bigr)
\right|
\right)
=
0.
\end{equation*}

Finally, we verify the stationarity condition with respect to
\(d_{\rvs,0}\). Since \(\mu_{\rvs}=0\), substituting
\(\omega_{\rvs}=W_{\rvs}^{\lambda(b)}(\mA(b))\) and
\(\beta_{\rvs}-\alpha_{\rvs}
=
K_{\rvs}^{\lambda(b)}(\mA(b))\) gives
\begin{align*}
\frac{\partial L}{\partial d_{\rvs,0}}
&=
\log(1-a_{\rvs,0})
+
W_{\rvs}^{\lambda(b)}
\bigl(\mA(b)\bigr)
-
\gamma
W_{n_1(\rvs)}^{\lambda(b)}
\bigl(\mA(b)\bigr)
+
p_{\rvs,0}
K_{\rvs}^{\lambda(b)}
\bigl(\mA(b)\bigr).
\end{align*}
Therefore, it remains to show that the above expression is equal to zero.

From the definition of \(\mW^\lambda(\mA)\), its $\rvs$-th component can be written as
\begin{align*}
W_{\rvs}^{\lambda}(\mA)
&=
H(a_{\rvs,0})
-
\lambda
\left|
a_{\rvs,0}-p_{\rvs,0}
\right|
+
\gamma
a_{\rvs,0}
W_{n_0(\rvs)}^{\lambda}(\mA)
+
\gamma
(1-a_{\rvs,0})
W_{n_1(\rvs)}^{\lambda}(\mA).
\end{align*}
Thus,
\begin{align*}
\frac{\partial L}{\partial d_{\rvs,0}}
&=
\log(1-a_{\rvs,0})
+
H(a_{\rvs,0})
-
\lambda(b)
\left|
a_{\rvs,0}-p_{\rvs,0}
\right|
\\
&\quad
+
\gamma
a_{\rvs,0}
\left(
W_{n_0(\rvs)}^{\lambda(b)}(\mA(b))
-
W_{n_1(\rvs)}^{\lambda(b)}(\mA(b))
\right)
+
p_{\rvs,0}
K_{\rvs}^{\lambda(b)}
\bigl(\mA(b)\bigr).
\end{align*}
From the definition of
\(K_{\rvs}^{\lambda(b)}(\mA(b))\), we have
\begin{equation*}
\gamma
\left(
W_{n_0(\rvs)}^{\lambda(b)}(\mA(b))
-
W_{n_1(\rvs)}^{\lambda(b)}(\mA(b))
\right)
=
\log\left(
\frac{a_{\rvs,0}}{1-a_{\rvs,0}}
\right)
-
K_{\rvs}^{\lambda(b)}
\bigl(\mA(b)\bigr).
\end{equation*}
Therefore,
\begin{align*}
\frac{\partial L}{\partial d_{\rvs,0}}
&=
\log(1-a_{\rvs,0})
+
H(a_{\rvs,0})
+
a_{\rvs,0}
\log\left(
\frac{a_{\rvs,0}}{1-a_{\rvs,0}}
\right)
\\
&
+
\left(
p_{\rvs,0}-a_{\rvs,0}
\right)
K_{\rvs}^{\lambda(b)}
\bigl(\mA(b)\bigr)
-
\lambda(b)
\left|
a_{\rvs,0}-p_{\rvs,0}
\right| \\
&=
\left(
p_{\rvs,0}-a_{\rvs,0}
\right)
K_{\rvs}^{\lambda(b)}
\bigl(\mA(b)\bigr)
-
\lambda(b)
\left|
a_{\rvs,0}-p_{\rvs,0}
\right|.
\end{align*}

For \(\rvs\neq s_1\), \(a_{\rvs,0}=p_{\rvs,0}\), and thus
\begin{equation*}
\frac{\partial L}{\partial d_{\rvs,0}}
=
0.
\end{equation*}
For \(\rvs=s_1\), using
\begin{equation*}
p_{s_1, 0}-a_{s_1, 0}
=
\sigma_{s_1}u_{s_1}(b), \text{ and }
K_{s_1}^{\lambda(b)}
\bigl(\mA(b)\bigr)
=
\sigma_{s_1}\lambda(b),
\end{equation*}
we obtain
\begin{align*}
\frac{\partial L}{\partial d_{s_1, 0}}
&=
\sigma_{s_1}u_{s_1}(b)
\left(
\sigma_{s_1}\lambda(b)
\right)
-
\lambda(b)u_{s_1}(b)
=
0.
\end{align*}

\end{enumerate}